\documentstyle[12pt,nomargid]{amsart}
\makeatletter
\def\eqnarray{\stepcounter{equation}\let\@currentlabel=\theequation
\global\@eqnswtrue
\global\@eqcnt\z@\tabskip\@centering\let\\=\@eqncr
$$\halign to \displaywidth\bgroup\@eqnsel\hskip\@centering
  $\displaystyle\tabskip\z@{##}$&\global\@eqcnt\@ne
  \hfil$\displaystyle{\hbox{}##\hbox{}}$\hfil
  &\global\@eqcnt\tw@ $\displaystyle\tabskip\z@
  {##}$\hfil\tabskip\@centering&\llap{##}\tabskip\z@\cr}
\@addtoreset{equation}{section}
  \def\theequation{\thesection.\arabic{equation}}

\def\endproof{\parfillskip 1.0em plus 1fil \lineskiplimit \z@
  \lineskip \z@ \par \@tempdima \prevdepth \advance \@tempdima 0.7em
  \ifdim \@tempdima > \baselineskip \vskip -\@tempdima \else
  \vskip -\baselineskip \fi \@tempdima \prevdepth
  \hbox to \hsize {\vrule height 0pt depth \@tempdima width 0pt \hfil
  $\Box$}\endtrivlist}
\makeatother
\def\mbar#1{\kern 0.1em\overline{\kern -0.1em #1 \kern -0.1em} \kern 0.1em}
\newtheorem{pro}{Proposition}[section]
\newtheorem{thm}{Theorem}[section]
\newtheorem{col}{Corollary}[section]
\newtheorem{lem}{Lemma}[section]
\def\Ad{\mathop{\rm Ad}\nolimits}
\def\ad{\mathop{\rm ad}\nolimits}
\def\Aut{\mathop{\rm Aut}\nolimits}
\def\dim{\mathop{\rm dim}\nolimits}
\def\id{\mathop{\rm id}\nolimits}
\def\Int{\mathop{\rm Int}\nolimits}
\def\Pr{\mathop{\rm pr}\nolimits}
\def\rank{\mathop{\rm rank}\nolimits}

\begin{document}

\title{Differential geometry of Toda systems}

\author{Alexander V. Razumov}

\address{Institute for High Energy Physics, 142284 Protvino, Moscow
region, Russia. \hbox{E-mail}: razumov@@m10.ihep.su}

\author{Mikhail V. Saveliev
}
\address{Laboratoire de Physique Th\'eorique de
l'Ecole Normale Sup\'erieure\footnote{Unit\'e Propre du Centre
National de la Recherche Scientifique, associ\'ee \`a
l'Ecole Normale Sup\'erieure et \`a l'Universit\'e de Paris-Sud.},
24 rue Lhomond, 75231 Paris C\'EDEX 05, ~France (On leave of absence from
Institute for High Energy Physics, 142284 Protvino, Moscow region,
Russia). E-mail: saveliev@@physique.ens.fr}

\maketitle

\begin{abstract}
In the present paper we give a differential geometry formulation
of the basic dynamical principle of the group--algebraic approach
\cite{LeS92} --- the grading condition --- in terms of some holomorphic
distributions on flag manifolds associated with the parabolic subgroups of
a complex Lie group; and a derivation of the corresponding nonlinear
integrable systems, and their general solutions.  Moreover, the reality
condition for these solutions is introduced.  For the case of the simple
Lie groups endowed with the canonical gradation, when the systems in
question are  reduced to the abelian Toda equations, we obtain the
generalised Pl\"ucker representation for the pseudo--metrics specified by
the K\"ahler metrics on the flag manifolds related to the maximal
nonsemisimple parabolic subgroups; and the generalised infinitesimal
Pl\"ucker formulas for the Ricci curvature tensors of these
pseudo--metrics.  In accordance with these formulas, the  fundamental
forms of the pseudo--metrics and the Ricci curvature tensors are expressed
directly in terms of the abelian Toda fields, which have here the sense of
K\"ahler potentials.
\end{abstract}

\section{Introduction}

The aim of the present paper is to describe some aspects of a differential
and algebraic geometry foundation of a wide class of nonlinear integrable
systems generated by a flat connection in the trivial fibre bundle $M
\times G \to M$, with a complex Lie group $G$ and a two--dimensional
manifold $M$ supplied with a complex structure. The principal example here
is the well--known finite nonperiodic Toda systems (abelian as well as the
nonabelian versions) associated with an arbitrary finite--dimensional
simple Lie group, and their affine deformations resulting in the periodic
Toda systems.

At present, there is the so--called group--algebraic approach \cite{LeS92}
which proved itself very efficient for a constructive investigation of the
integrability problem for nonlinear systems in low dimensions. In the
framework of this method, the integrability criteria is related to such
properties of the internal symmetry group of the equations under
consideration, as its solvability and finiteness of the growth; and in turn
is realised in the relevant properties of the characteristic algebra of
the corresponding dynamical system \cite{LSS82,LeSh82}, whose generators
can be written, speaking in modern terms, as a $W$--algebra elements. At
the same time, a differential and algebraic geometry setting of nonlinear
integrable systems, which seems to be quite important, however, is not
clear enough yet.

Our present paper is mainly devoted to the extrinsic geometry of
integrable systems. Here we give a differential geometry formulation of
the basic dynamical principle of the group--algebraic approach
\cite{LeS92} --- the grading condition, and its realisation in terms of
some holomorphic distributions on flag manifolds $F_{\pm}=G/\tilde
B_{\mp}$, where $\tilde B_{\mp }$ are the parabolic subgroups of $G$
associated with a given ${\Bbb Z}$--gradation of the Lie algebra ${\goth
g}$ of the Lie group $G$.  Such distributions have already been studied in
mathematical literature as superhorizontal distributions \cite{BuR90}.
Moreover, the consideration of the nonlinear Toda type systems given in
our paper seems to be closely related to variations of the Hodge
structures in the spirit of Ph.\ A. Griffiths \cite{Gri70,Sim92}.
Further, we consider the derivation of the related nonlinear systems. In
particular, for the simple Lie groups one comes to the Toda systems, and
just for them we reproduce a construction of their general solutions
already known from
\cite{LeS92}, where they were obtained using purely Lie group methods.
Moreover, here we introduce the reality condition for these solutions in
terms of the corresponding mappings $M\to F_{\pm}$.  Finally, for the case
of the complex simple Lie algebra endowed with the canonical gradation,
when the systems under consideration are reduced to the equations of the
abelian Toda system, we derive the generalised Pl\"ucker representation
for the pseudo--metrics on $M$ specified by the K\"ahler metrics on the
flag manifolds related to the maximal nonsemisimple parabolic subalgebras
(closely related to the fundamental representations of ${\goth g}$); and
the generalised infinitesimal Pl\"ucker formulas for the Ricci curvature
tensors of these pseudo--metrics.  In accordance with these formulas, the
fundamental forms of the pseudo--metrics on $M$ and the Ricci curvature
tensors are expressed directly in terms of the abelian Toda fields, which
have here the sense of K\"ahler potentials.

Note that the fact that for the case of the canonical distributions the
Ricci curvature tensors are connected with the corresponding fundamental
forms of the pseudo--metrics has been conjectured in \cite{Giv89} and then
proved in \cite{Pos91}, see also \cite{Yo87} for the symplectic algebra,
while in a very brief form, and without any relation to integrable
systems. Moreover, this generalisation of the Pl\"ucker representation and
formulas has been obtained in \cite{GS93} by an explicit calculation of
the K\"ahler potentials on the corresponding manifolds associated with the
fundamental representations of the classical simple Lie algebras (series
$A$, $B$, $C$, $D$) endowed with the canonical gradation. This last paper
comes back to the pioneering, and also quite recent investigation
\cite{GM92,GM93} of some geometrical structures behind $W$--algebra for the
case of the abelian $A$--Toda system  as a notion of a $W_{A}$--geometry of the
complex projective target manifolds associated to this system. In fact, just
the very remarkable papers \cite{GM92,GM93}, together with numerous discussions
and contacts with J.--L. Gervais, both illuminating and pleasant, served as a
vigorous stimulus for our present study of the generalised Pl\"ucker
relations.

It seems to us very believable that the relevant modification of the last
part of our present paper, concerning generalised Pl\"ucker representation
and formulas, can be done for a more wide class of nonlinear integrable
partial differential equations, in particular for nonabelian versions of
the finite nonperiodic Toda system.  Moreover, all what we have said up to
now, mainly concerns with the finite--dimensional manifolds associated
with finite--dimensional Lie algebras, and in turn with the finite
systems. There is also a number of very interesting papers where the
infinite--dimensional K\"ahler geometry and Grassmannians associated with
the group of smooth based loops on a connected compact Lie group were
investigated, see e.g. \cite{Ka85} and references therein; and some
aspects of the differential geometry of the affine Toda systems are
treated, see e.g. \cite{Bu92,Bu93,Wo93,BPW93}.  Here the arising
(flag) manifolds behave in many respects like finite--dimensional ones.
At the same time, a study of the infinite--dimensional K\"ahler manifolds
associated with infinite--dimensional  Lie algebras, in particular,
algebras of the diffeomorphisms groups, should deal with the notions and
objects of a novel nature, see e.g. a consideration of the
$W_{\infty}$-geometry of the continuous $A_{\infty}$-Toda system in
\cite{SS93}. However, the $W$-geometry of the dynamical systems related to
the infinite--dimensional Lie algebras represents much more complicated
problem, and we will not discuss it here at all.

Concluding the Introduction, we would like to explain why we supply our
paper with the appendices containing more or less known information
concerning the geometry of the complex and real Lie groups; the parabolic
subalgebras of the simple Lie algebras and the associated flag manifolds;
and ${\Bbb Z}$--gradations. The point is that at present many
theoreticians work with the problems related to the Toda systems, while
our language is mainly differential and algebraic geometry one. Since, as
far as we know, it is not commonly accepted by them in a whole extent, we
give there some notations and definitions to simplify reading of our
paper.  Moreover, some definitions used in the main body of the paper,
e.g. a relation of ${\goth g}$--valued 1--forms and a connection forms,
vertical automorphism group and gauge group, are figured in physical and
mathematical literature in different forms and, sometimes, with different
meanings; some notions are defined on a different level of generality,
e.g. matrix--valued and generic 1-forms taking values in a Lie algebra;
this is why we found it possible and reasonable to give some definitions
which synthesize and interpolate, in a sense, between them.

\section{Zero Curvature Representation of Toda Equations}

\subsection{Flat Connections}

Let $M$ be a two--dimensional manifold supplied with a complex structure
$J^M$, so that $M$ can be considered as a one--dimensional complex
manifold. Let $G$ be a complex semisimple Lie group, and ${\goth g}$ be
the corresponding complex Lie algebra (see appendix A). Consider a ${\goth
g}$--valued 1--form $\omega$ on $M$. Note that any such a form corresponds
to a connection form of some connection in the trivial holomorphic
principal fibre bundle $M \times G \to M$, so we call $\omega$ also a
connection form, or simply a connection. Suppose that the form $\omega$
satisfies the condition
\begin{equation}
d\omega(X, Y) + [\omega(X), \omega(Y)] = 0 \label{2.1}
\end{equation}
for any vector fields $X$, $Y$ on $M$.  In this case the corresponding
connection in $M \times G \to M$ has zero curvature, in other words it is
a flat connection. Hence, it is natural to call relation
(\ref{2.1}) the zero curvature condition. Using a local coordinate $z$ in
$M$, one can write
\begin{equation}
\omega = \omega_- dz + \omega_+ d\bar z, \label{2.2}
\end{equation}
where $\omega_{\pm}$ are some mappings from $M$ to ${\goth g}$. In what
follows the superscripts $-$ and $+$ mean for 1--forms on $M$ the
corresponding components in the expansion over the local basis formed by $dz$
and $d\bar z$. In terms of $\omega_{\pm}$ the zero curvature condition
takes the form
\begin{equation}
\partial_z \omega_+  - \partial_{\bar z} \omega_- + [\omega_-, \omega_+] =
0. \label{2.3}
\end{equation}
Here and below we use the notation
\begin{equation}
\partial_z \equiv \partial/\partial z, \qquad \partial_{\bar z} =
\partial/\partial \mbar z. \label{2.3a}
\end{equation}
Choosing a basis in ${\goth g}$ and considering the components of
the expansion of $\omega_{\pm}$ over this basis as fields, we can treat
the zero curvature condition as a nonlinear system of partial differential
equations for the fields.

Note that the group of vertical automorphisms of the fiber bundle $M\times
G \to M$ is parametrised by a mapping from $M$ to $G$. According to the
tradition accepted in the physical literature,
we call this group the group of gauge transformations.
Let $\psi$ be a mapping from $M$ to $G$, and $\omega$ be a connection
form; the gauge transformed connection form $\omega^\psi$ is given by the
relation
\begin{equation}
\omega^\psi(x) = \Ad (\psi^{-1}(p)) \omega(x) + \psi^*
\theta(x) \label{2.4}
\end{equation}
for any $p \in M$ and $x \in T^{\Bbb C}_p(M)$.  The notation $\psi^{-1}$
in (\ref{2.4}) means the mapping connected with the mapping $\psi$ by
\begin{equation}
\psi^{-1}(p) \equiv (\psi(p))^{-1}. \label{2.5}
\end{equation}
Relation (\ref{2.4}) can be written formally as
\begin{equation}
\omega^\psi = \Ad (\psi^{-1}) \circ \omega + \psi^* \theta ,\label{2.6}
\end{equation}
where $\circ$ means the composition of the mappings.  For the case of a
linear group, $\omega$ is a matrix valued 1--form, $\psi$ is a matrix
valued function, and Eq.~(\ref{2.6}) takes the form
\begin{equation}
\omega^\psi = \psi^{-1} \omega \psi + \psi^{-1} d \psi.
\end{equation}

The zero curvature condition is invariant with respect to the gauge
transformations (\ref{2.6}). In other words, if a connection $\omega$
satisfies this condition, then the connection $\omega^\psi$ also satisfies
it. It is convenient to call the gauge transformations defined by
Eq.~(\ref{2.6}), $G$--gauge transformations.

Suppose that the manifold $M$ is simply connected, then for any flat
connection $\omega$ there exists a mapping $\varphi$ from $M$ to $G$ such
that
\begin{equation}
\omega = \varphi^* \theta, \label{2.7}
\end{equation}
where $\theta$ is the holomorphic Maurer--Cartan form of the group $G$
(see appendix A). Note that for a linear group one can write
\begin{equation}
\omega = \varphi^{-1} d \varphi.
\end{equation}
{}From the other hand, for any mapping $\varphi$ the connection
${}^\varphi\omega$ given by
\begin{equation}
{}^\varphi\omega = \varphi^* \theta, \label{2.7a}
\end{equation}
satisfies Eq.~(\ref{2.1}). Moreover, ${}^\varphi \omega = 0$ if and
only if $\varphi$ is a constant mapping.

For any two mappings $\varphi, \psi: M \to G$ we define a new mapping
$\varphi \psi$ by the relation
\begin{equation}
\varphi \psi(p) \equiv \varphi(p)\psi(p). \label{2.8}
\end{equation}

\begin{pro} \label{p:2.1}
For any two mappings $\varphi, \psi: M \to G$, the relation
\begin{equation}
{}^{\varphi  \psi} \omega = \Ad (\psi^{-1}){}^\varphi \omega +
{}^\psi \omega \label{2.9}
\end{equation}
is valid.
\end{pro}
\begin{pf}
For any $p \in M$ and $x \in T^{\Bbb C}_p(M)$ we have
\begin{equation}
{}^{\varphi  \psi} \omega(x) = (\varphi  \psi)^* \theta(x) =
\theta((\varphi  \psi)_{*p}(x)). \label{2.10}
\end{equation}
It is not difficult to show that
\begin{equation}
(\varphi  \psi)_{*p} = R_{\psi(p)*\varphi(p)} \circ \varphi_{*p} +
L_{\varphi(p)*\psi(p)} \circ \psi_{*p}. \label{2.11}
\end{equation}
Now using Eqs.~(\ref{a.5a}) and (\ref{a.5b}), we arrive at Eq.~(\ref{2.9}).
\end{pf}

\begin{col} \label{c:2.1}
For any two mappings $\varphi, \psi: M \to G$,
\begin{equation}
({}^\varphi\omega)^\psi = {}^{\varphi  \psi}\omega, \qquad
{}^{\varphi^{-1}} \omega = - \Ad (\varphi)({}^\varphi \omega).
\end{equation}
\end{col}

\begin{pro} \label{p:2.2}
If the equality
\begin{equation}
{}^{\varphi} \omega = {}^{\varphi'} \omega, \label{2.13}
\end{equation}
is valid, then $\varphi'  \varphi^{-1}$ is a constant mapping.
\end{pro}
\begin{pf}
Performing the gauge transformation corresponding to the mapping
$\varphi^{-1}$, from Eq.~(\ref{2.13}) we get the equality
\begin{equation}
{}^{\varphi'  \varphi^{-1}} \omega = 0; \label{2.14}
\end{equation}
hence, $\varphi'  \varphi^{-1}$ is a constant mapping.
\end{pf}

Therefore, in the case under consideration, any flat connection can be
gauge transformed to zero. In this sense system (\ref{2.3}) is trivial.
{}From the other hand, the majority of two--dimensional integrable equations
can be obtained from system (\ref{2.3}) by imposing some
gauge--noninvariant constraints on the connection $\omega$. Note that, in
general, for the case of infinite dimensional Lie algebras and Lie groups
one needs a generalisation of the scheme, see Refs.~\cite{LeS92,OSU93},
but in the present paper we restrict ourself to the finite dimensional
case.  Consider one of the methods to impose the conditions in question,
giving, in fact, a differential--geometric formulation of the
group--algebraic approach for integrating nonlinear systems in the spirit
of Ref.~\cite{LeS92}.

\subsection{Grading Condition}

Suppose, following Ref.~\cite{LeS92}, that the Lie algebra ${\goth g}$ is
a ${\Bbb Z}$--graded Lie algebra, (see appendix B). The first condition we
impose on the connection $\omega$ is the following. Let $\tilde {\goth
b}_\pm$ be the subalgebras of ${\goth g}$ defined by Eq.~(\ref{b.24}).
Require that the (1,0)--component of the form $\omega$ takes values in
${\goth b}_-$, and its (0,1)--component takes values in ${\goth b}_+$.  We
call this condition the grading condition.  Any connection $\omega$
satisfying the grading condition is of the form ${}^\varphi \omega$ for
some mapping $\varphi : M \to G$; however not any mapping $\varphi$ leads
to the connection ${}^\varphi\omega$ satisfying this condition. Let us
formulate requirements which should be imposed on $\varphi$ to guarantee
the validity of the grading condition.

Denote by $\tilde B_\pm$ the connected subgroups of $G$ corresponding to
the subalgebras $\tilde {\goth b}_\pm$. Since $\tilde {\goth b}_\pm$ are
parabolic subalgebras of ${\goth g}$ (see appendix B), the subgroups
$\tilde B_\pm$ are parabolic subgroups of $G$; hence, the homogeneous
spaces $F_\pm = G/\tilde B_\mp$ are flag manifolds.  Let $\pi_{\pm}: G \to
F_{\pm}$ be the canonical projections. Define the mappings $\varphi_{\pm}
: M \to F_{\pm}$ by
\begin{equation}
\varphi_{\pm} = \pi_{\pm} \circ \varphi. \label{2.17}
\end{equation}

\begin{thm} \label{t:2.1}
The connection ${}^\varphi\omega$ satisfies the grading conditions if and
only if the mapping $\varphi_-$ is holomorphic, and the mapping
$\varphi_+$ is antiholomorphic.
\end{thm}

\begin{pf}
Suppose that the 1-form
$({}^\varphi \omega)^{(0,1)}$ takes values in ${\goth b}_+$.  Thus, for any
$x \in T^{\Bbb C}_p(M)$ we have
\begin{equation}
(\pi_{-*e} \circ {}^\varphi \omega(\bar P^M_p(x)) = 0, \label{2.18}
\end{equation}
where the mapping $\bar P^M_p$ projects the tangent vector $x$ to its
$(0,1)$--component.
Equality (\ref{2.18}) can be rewritten as
\begin{equation}
\theta((\varphi_{-*p} \circ \bar P^{M}_p)(x)) = 0. \label{2.19}
\end{equation}
Note that, since $\theta$ is a form of the type (1,0), the equality
\begin{equation}
\theta(x) = 0, \qquad x \in T^{\Bbb C}_g(G), \label{2.20}
\end{equation}
is valid if and only if
\begin{equation}
P^G_g(x) = 0 \label{2.21}
\end{equation}
(see appendix A). Hence, from Eq.~(\ref{2.19}) we have
\begin{equation}
P^G_{\varphi_-(p)} \circ \varphi_{-*p} \circ \bar P^M_p = 0
\label{2.22}
\end{equation}
for any point $p \in M$. The mapping $\varphi_{+*p}$ is real; thus,
after the complex conjugation of Eq.~(\ref{2.22}), we get
\begin{equation}
\bar P^G_{\varphi_-(p)} \circ \varphi_{-*p} \circ P^M_p = 0.
\label{2.23}
\end{equation}
It follows from Eqs.~(\ref{2.22}) and (\ref{2.23}) that
\begin{equation}
J^G_{\varphi_-(p)} \circ \varphi_{-*p} = \varphi_{-*p} \circ J^{M}_p,
\label{2.24}
\end{equation}
and so the mapping $\varphi_-$ is holomorphic.

Suppose now that the mapping $\varphi_-$ is holomorphic. Reversing the
arguments given above, we conclude that the form $\omega^{(0,1)}$ takes
values in ${\goth b_+}$.

The case of the mapping $\varphi_+$ can be considered in the same way.
\end{pf}

We call a mapping $\varphi$ generating a connection satisfying the grading
condition, a mapping satisfying the grading condition. Now, following
again Ref.~\cite{LeS92}, perform a further specification of the grading
condition. Here we confine ourselves by a special choice of such a
specification leading to the Toda type systems; for a more general case
see Ref.~\cite{LeS92}.

Let us require that the (1,0)--component of the connection $\omega$ takes
values in the linear space ${\goth g}_{-1} \oplus {\goth g}_0$, and the
(0,1)--component of it takes values in the linear space ${\goth g}_0
\oplus {\goth g}_{+1}$. To reformulate the grading condition as a condition
imposed on the corresponding mapping $\varphi$, we introduce some
holomorphic distributions on the manifolds $F_+$ and $F_-$. To this end,
define subspaces $\tilde {\goth m}_\pm$ of ${\goth g}$ by
\begin{equation}
\tilde {\goth m}_+ \equiv \bigoplus_{m \le 1} {\goth g}_m, \qquad \tilde
{\goth m}_- \equiv \bigoplus_{m \ge -1} {\goth g}_m.
\end{equation}
According to the convention of appendix B, in the case of the canonical
gradation we use the notations without tildes. Note that in this case we
can also define the subspaces ${\goth m}_\pm$ as
\begin{equation}
{\goth m}_\pm \equiv \{ x \in {\goth g} \mid [x, [{\goth b}_\mp, {\goth
b}_\mp]] \in {\goth b}_\mp \}.
\end{equation}

The subspace $\tilde {\goth m}_+$ is invariant with respect to the
adjoint  action of the subgroup $\tilde B_-$ in ${\goth g}$. Let $p \in
F_+$, and $g$ be any element of $G$ such that $\pi_+(g) = p$. Define the
subspace ${\cal M}_{+p} \subset T^{(1,0)}_p(F_+)$ by
\begin{equation}
{\cal M}_{+p} \equiv \pi_{+*}(\tilde {\goth m}_{+g}),
\end{equation}
where
\begin{equation}
\tilde {\goth m}_{+g} \equiv L_{g*}(\tilde {\goth m}_+).
\end{equation}
The subspaces ${\cal M}_{+p}$, $p \in F_+$, generate a distribution on $F_+$
which will be denoted ${\cal M}_+$. In the same way we can define an
analogous distribution ${\cal M}_-$ on $F_-$.

\begin{thm} \label{t:2.2}
The connection ${}^\varphi \omega$ satisfies the grading condition if an
only if the mapping $\varphi_-$ is holomorphic and
$\varphi_{-*}(\partial_{z p}) \in {\cal M}_{-\varphi_-(p)}$ for any $p \in
M$, while the mapping $\varphi_+$ is antiholomorphic and
$\varphi_{+*}(\partial_{\bar z p}) \in {\cal M}_{+\varphi_+(p)}$ for any
$p \in M$.
\end{thm}

\begin{pf}
The proof is similar to the proof of Theorem \ref{t:2.1}.
\end{pf}

It is natural in the situation, described in theorem \ref{t:2.2}, to say
that the mapping $\varphi_+$ is tangent to the distribution ${\cal M}_+$,
while the mapping $\varphi_-$ is tangent to the distribution ${\cal M}_-$.
Note that this condition is directly related to the notion of a
superhorizontal mapping \cite{BuR90}.

It is clear that the grading condition is not invariant under the action
of an arbitrary $G$--gauge transformation, but it is invariant under the
action of gauge transformations (\ref{2.6}) generated by the mappings
taking values in the subgroup $\tilde H$, corresponding to the subalgebra
$\tilde {\goth h} \equiv {\goth g}_0$. Such transformations form a
subgroup of the group of $G$--gauge transformations. Below we call a gauge
transformation from this subgroup an $\tilde H$--gauge transformation. To
get a system of equations having not $G$--gauge invariance, we should impose
further restrictions on the connection form. In fact, we choose following
Ref.~\cite{LeS92} another way leading to the system in question. It
consists in constructing some $\tilde H$--gauge invariant quantities,
and in rewriting the zero curvature conditions in terms of them. To this
end, let us first consider the structure of the holomorphic principal fibre
bundles $G \to F_\pm$.

\subsection{Modified Gauss Decomposition}

Recall that the subspaces $\tilde {\goth n}_\pm$ defined by Eq.~(\ref{b.22})
are subalgebras of ${\goth g}$. Denote by $\tilde N_{\pm}$ the corresponding
connected subgroups of $G$. Let us formulate and prove two useful lemmas.

\begin{lem} \label{l:3.1}
Let $h$ be an arbitrary element of $G$; if the set $\tilde N_+ \cap
h\tilde B_- \ne \emptyset$, it contains just one element of $G$.
\end{lem}

\begin{pf}
Note that $\tilde N_+ \cap \tilde B_- = \{e\}$. Suppose that $n_1, n_2 \in
\tilde N_+ \cap g \tilde B_-$, then there exist the elements $b_1, b_2 \in
\tilde B_-$, such that
\begin{equation}
n_1 = h b_1, \qquad n_2 = h b_2. \label{2.26}
\end{equation}
This relation implies the equality
\begin{equation}
n_1^{-1} n_2 = b_1^{-1} b_2. \label{2.27}
\end{equation}
Hence, we have $n_1^{-1} n_2 = e$. That was to be proved.
\end{pf}

\begin{lem} \label{l:3.2}
The set $\pi_+(\tilde N_+)$ is an open set.
\end{lem}

\begin{pf}
It is clear that $\pi_+(\tilde N_+) = \pi_+(\tilde N_+ \tilde B_-)$. The
subspaces $\tilde {\goth n}_+$ and $\tilde {\goth b}_-$ are such
subalgebras of ${\goth g}$ that
\begin{equation}
{\goth g} = \tilde {\goth n}_+ \oplus \tilde {\goth b}_-; \label{2.28}
\end{equation}
hence, the mapping $(n, b) \in \tilde N_+ \times \tilde B_- \to nb \in G$
is regular for any $(n, b) \in \tilde N_+ \times \tilde B_-$ (see, for
example, \cite[Ch.  VI]{Hel62}). It follows from this property that the
set $\tilde N_+ \tilde B_-$ is open.  From the other hand, the mapping
$\pi_+$ is an open mapping; thus $\pi_+(\tilde N_+)$ is an open set.
\end{pf}

Using the assertions of these two lemmas, we can define a local
section $s_+$ of the fibre bundle $G \to F_+$, assuming that
\begin{equation}
s_+(p) \equiv \tilde N_+ \cap (\pi_+)^{-1}(p), \qquad p \in \pi_+(\tilde
N_+). \label{2.29}
\end{equation}
Since $\tilde N_+$ is a holomorphic submanifold of $G$, the section $s_+$ is
holomorphic.

The subspaces $\tilde {\goth n}_{+g} \equiv L_{g*}(\tilde {\goth n}_+)$,
$g \in G$, generate an involutive holomorphic distribution ${\cal N}_+$ on
$G$. The image $\tilde N_+$ of the section $s_+$ is an integral manifold
of this distribution. Note that for any $g \in G$ the set $g \tilde N_+$
is also an integral manifold of the distribution ${\cal N}_+$. The
following proposition is now almost evident.

\begin{pro} \label{p:2.5}
There exists an open covering $\{U_{+a}\}_{a \in A}$ of the manifold $F_+$,
and a family of local holomorphic sections $\{s_{+a}\}_{a \in A}:
U_{+a} \to G$ of the fibre bundle $G \to F_+$, such that for any $a
\in A$ the set $s_{+a}(U_{+a})$ is an integral manifold of the distribution
${\cal N}_+$.
\end{pro}

\begin{pf}
As the first element of the required covering and the corresponding
section we can take the set $U_+ = \pi_+(\tilde N_+)$ and the section
$s_+$ defined above. Let $p \not\in U_+$, and $g \in (\pi_+)^{-1}(p)$.
The set $g \tilde N_+$ possesses properties similar to the properties of
the set $\tilde N_+$.  Namely, if $g\tilde N_+ \cap  h \tilde B_- \ne
\emptyset$, then the set $g \tilde N_+ \cap h \tilde B_-$ contains just
one point, and the set $\pi_+(g \tilde N_+)$ is open. Therefore, we can
define a local holomorphic section $s'_+: U'_+ \to G$, where $U'_+ =
\pi^+(g \tilde N^+)$. Here the set $U'_+$ contains the point $p$.
Repeating this procedure, we get a family of local holomorphic sections of
the fibre bundle $G \to F^+$ with the required properties.
\end{pf}

Actually we shall consider families of local sections constructed with the
help of the procedure used in the proof of Proposition \ref{p:2.6}. In
this case, for any $a \in A$ we have
\begin{equation}
s_{+a}(U_{+a}) = g_{+a} \tilde N_+ \equiv \tilde N_{+a}
\end{equation}
for some $g_{+a} \in G$. It is clear that a similar family of local
sections can be constructed also for the fibre bundle $G \to F_-$. If the
Lie algebra ${\goth g}$ is endowed with an involutive antilinear
automorphism consistent with the ${\Bbb Z}$--gradation, such a family of
sections of the fibre bundle $G \to F_-$ can be constructed on the base of
the given family of sections of the fibre bundle $G \to F_+$. The
corresponding method to do that is considered in section 3.

It is known that any family of holomorphic local sections of a holomorphic
principal fibre bundle, covering the base space, allows to introduce an
atlas of the fibre bundle under consideration. The corresponding procedure
in our case looks as follows.

Let $ a\in A$, consider a holomorphic mapping $m_{+a}:
(\pi_+)^{-1}(U_{+a}) \to G$ defined as
\begin{equation}
m_{+a}(g) \equiv s_{+a}(\pi_+(g)), \qquad g \in (\pi_+)^{-1}(U_{+a}).
\label{2.71}
\end{equation}
This mapping allows to introduce another holomorphic mapping $b_{-a}$
defined on $(\pi_+)^{-1}(U_{+a})$ by
\begin{equation}
b_{-a}(g) \equiv (m_{+a}(g))^{-1} g. \label{2.72}
\end{equation}
Thus, for any $g \in (\pi_+)^{-1}(U_{+a})$ we can write
\begin{equation}
g = m_{+a}(g) b_{-a}(g). \label{2.73}
\end{equation}
Since $s_{+a}$ is a section of the fibre bundle $G \to F_+$, i.e.,
\begin{equation}
\pi_+ \circ s_{+a} = \id_{U_{+a}}, \label{2.74}
\end{equation}
we have
\begin{equation}
\pi_+ (m_{+a}(g)) = \pi_+(g), \label{2.75}
\end{equation}
and from (\ref{2.73}) it is readily seen that the mapping $b_{-a}$ takes
values in the subgroup $\tilde B_-$. Note here that the mappings $m_{+a}$ and
$b_{-a}$ have the following properties:
\begin{equation}
m_{+a}(gb) = m_{+a}(g), \qquad b_{-a}(gb) = b_{-a}(g)b \label{2.76}
\end{equation}
for any $b \in B_-$.

It is clear that the mapping $\psi_{+a}: (\pi_+)^{-1}(U_{+a}) \to U_{+a}
\times \tilde B_-$, defined as
\begin{equation}
\psi_{+a}(g) \equiv (\pi_+(g), b_{-a}(g)), \label{2.77}
\end{equation}
provides a local trivialisation of the fibre bundle $G \to F_+$.
Considering all possible values of the index $a$, we get an atlas of this
fibre bundle.

Let $g$ be such an element of $G$ that $\pi_+(g) \in U_{+a} \cap U_{+b}$.
In this case, using Eq.~(\ref{2.76}), we can write
\begin{equation}
b_{-a}(g) = b_{-a}(s_{+b}(p) b_{-b}(g)) = b_{-a}(s_{+b}(p))
b_{-b}(g), \label{2.78}
\end{equation}
where $p = \pi_+(g)$. Hence we have
\begin{equation}
b_{-a}(g) = b_{-ab}(\pi_+(g)) b_{-b}(g), \label{2.78a}
\end{equation}
where
\begin{equation}
b_{-ab} \equiv b_{-a} \circ s_{+b}. \label{2.79}
\end{equation}
It is clear that the mappings $b_{-ab}$, $a, b \in A$, are the transition
functions of the atlas we have defined. These transition functions are
evidently holomorphic.

Let again $\pi_+(g) \in U_{+a} \cap U_{+b}$. In this case we have
\begin{equation}
g = m_{+a}(g) b_{-a} = m_{+b}(g) b_{-b}(g). \label{2.79a}
\end{equation}
This relation, with the account of Eq.~(\ref{2.78a}) gives
\begin{equation}
m_{+b}(g) = m_{+a}(g) b_{-ab}(\pi_+(g)). \label{2.79b}
\end{equation}

The groups $\tilde B_\pm$ have the holomorphic decomposition
\begin{equation}
\tilde B_\pm = \tilde N_\pm \tilde H. \label{2.85}
\end{equation}
This is true, in particular, for the case of the canonical gradation.
Thus, in our case we can uniquely represent the mapping $b_{-a}$ in the
form
\begin{equation}
b_{-a} = n_{-a}  h_{-a}, \label{2.80}
\end{equation}
where $n_{-a}$ and $h_{-a}$ are holomorphic mappings from
$(\pi_+)^{-1}(U_{+a})$ to the subgroups $\tilde N_-$ and $\tilde H$,
respectively.  Analogously for the transition functions $b_{-ab}$ we have
\begin{equation}
b_{-ab} = n_{-ab}  h_{-ab}, \label{2.81}
\end{equation}
where $n_{-ab}$ and $h_{-ab}$ are holomorphic mappings from $U_{+a} \cap
U_{+b}$ to $\tilde N_-$ and $\tilde H$, respectively.

{}From Eq.~(\ref{2.76}) we get the following relations
\begin{eqnarray}
&m_{+a}(gn) = m_{+a}(g), \qquad m_{+a}(gh) = m_{+a}(g),& \label{2.82} \\
&n_{-a}(gn) = n_{-a}(g) h_{-a}(g) n (h_{-a}(g))^{-1}, \qquad n_{-a}(gh) =
n_{-a}(g),& \label{2.83} \\
&h_{-a}(gn) = h_{-a}(g), \qquad h_{-a}(gh) = h_{-a}(g) h,& \label{2.84}
\end{eqnarray}
which are valid for any $n \in \tilde N_-$ and $\tilde h \in H$.

Using Eq.~(\ref{2.78a}), for any $g \in G$ such that $p = \pi_+(g) \in
U_{+a} \cap U_{+b}$, we also obtain
\begin{eqnarray}
&n_{-a}(g) = n_{-ab}(p) h_{-ab}(p) n_{-b}(g) (h_{-ab}(p))^{-1},&
\label{2.84a} \\
&h_{-a} = h_{-ab}(p) h_{-b}(g). \label{2.84b} &
\end{eqnarray}

\begin{pro} \label{p:2.6}
Any element $g \in (\pi_+)^{-1}(U_{+a}) = \tilde N_{+a} \tilde B_-$ can be
uniquely represented in the form
\begin{equation}
g = m_+ n_- h_-, \label{2.86}
\end{equation}
where $m_+ \in \tilde N_{+a},$ $n_- \in \tilde N_-$, $h_- \in \tilde H$.
The elements $m_+$, $n_-$, and $h_-$ holomorphically depend on $g$.
\end{pro}

\begin{pf}
We come to the representation (\ref{2.86}) putting
\begin{equation}
m_+ = m_{+a}(g), \qquad n_- = n_{-a}(g), \qquad h_- = h_{-a}(g).
\label{2.90}
\end{equation}
The uniqueness of the decomposition (\ref{2.86}) follows directly from the
fact that $\tilde N_+ \cap \tilde N_- = \tilde N_+ \cap \tilde H = \tilde
N_- \cap \tilde H = \{e\}.$
\end{pf}

There takes place the proposition similar to the previous one.

\begin{pro} \label{p:2.7}
Any element $g \in (\pi_-)^{-1}(U_{-a}) = \tilde N_{-a} \tilde B_+$ can be
uniquely represented in the form
\begin{equation}
g = m_- n_+ h_+, \label{2.91}
\end{equation}
where $m_- \in \tilde N_{-a},$ $n_+ \in \tilde N_+$, $h_+ \in \tilde H$.
The elements $m_-$, $n_+$, and $h_+$ holomorphically depend on $g$.
\end{pro}

We call decompositions (\ref{2.86}) and (\ref{2.91}) the {\it modified Gauss
decompositions}.

\subsection{Toda Equations}

Now we shall use the modified Gauss decomposition to define the needed
$\tilde H$--gauge invariant quantities, and to derive the equations they
satisfy.

\begin{pro} \label{p:2.8}
Let $\varphi : M \to G$ be an arbitrary mapping, and $p \in M$.

(i) There exists an open neighborhood $V_+$ of the point $p$, such that
the mapping $\varphi$ restricted to $V_+$ has a unique decomposition
\begin{equation}
\varphi = \mu_+  \nu_-  \eta_-, \label{2.92}
\end{equation}
where the mapping $\mu_+$ takes values in $\tilde N_{+a}$ for some $a \in
A$, while the mappings $\nu_-$ and $\eta_-$ take values in $\tilde N_-$
and $\tilde H$, respectively.

(ii) There exists an open neighborhood $V_-$ of the point $p$, such that
the mapping $\varphi$ restricted to $V_-$ has a unique decomposition
\begin{equation}
\varphi = \mu_-  \nu_+  \eta_+, \label{2.93}
\end{equation}
where the mapping $\mu_-$ takes values in $\tilde N_{-a}$ for some $a \in
A$, while the mappings $\nu_+$ and $\eta_+$ take values in $\tilde N_+$
and $\tilde H$, respectively.
\end{pro}

\begin{pf}
The proof is based on the modified Gauss decomposition (\ref{2.86}). It
is clear that we can find $a \in A$ such that $\varphi(p) \in
(\pi_+)^{-1}(U_{+a})$. Define a mapping $\mu_+$ by
\begin{equation}
\mu_+ \equiv m_{+a} \circ \varphi = s_{+a} \circ \pi_+ \circ \varphi =
s_{+a} \circ \varphi_+. \label{2.94}
\end{equation}
The domain of the mapping $\mu_+$ is the open set  $V_+
\equiv \varphi_+^{-1} (U_{+a})$. Introducing now the mappings
\begin{equation}
\nu_- \equiv n_{-a} \circ \varphi, \qquad \eta_- \equiv h_{-a} \circ
\varphi  \label{2.95}
\end{equation}
with the same domain, we come to the required decomposition
(\ref{2.92}).

The second part of the proposition can be proved in the similar way.
\end{pf}

\begin{col} \label{c:2.2}
For any $p \in M$, there exist an open set $V$, such that $p \in V$, and the
mapping $\varphi$ restricted to $V$ possesses simultaneously
decompositions {\rm (\ref{2.92})} and {\rm (\ref{2.93})}.
\end{col}

\begin{pro} \label{p:2.9}
If the mapping $\varphi: M \to G$ satisfies the grading condition, then the
mapping $\mu_-$ is holomorphic and the holomorphic 1--form ${}^{\mu_-}
\omega$ takes values in ${\goth g}_{-1}$; while the mapping $\mu_+$ is
antiholomorphic and the antiholomorphic 1--form ${}^{\mu_+} \omega$ takes
values in ${\goth g}_{+1}$.
\end{pro}

\begin{pf}
The required properties of the mappings $\mu_\pm$ follow directly from the
grading condition.
\end{pf}


Let $x_\pm$ be some fixed nonzero elements of ${\goth g}_{\pm 1}$. In the
case when we consider the ${\Bbb Z}$--gradation of ${\goth g}$ associated
with an integral embedding of $\mbox{\goth sl}(2, \Bbb C)$ into ${\goth
g}$, we can take as $x_\pm$ the corresponding elements defined by this
embedding.  Let ${\cal O}_\pm$ be the orbits of the restriction of the
adjoint action of the group $G$ to the subgroup $\tilde H$, generated by
the elements $x_\pm$. Note that the orbits ${\cal O}_\pm$ have the
following property. If the element $x$ belongs to ${\cal O}_+$ (${\cal
O}_-$) then for any nonzero $c \in {\Bbb C}$ the element $cx$ also
belongs to ${\cal O}_+$ (${\cal O}_-$). This follows from the fact that the
gradation operator $q$ (see Appendix B) generates similarity
transformations of the subspaces ${\goth g}_{\pm 1}$.

Denoting by $\tilde H_\pm$ the isotropy subgroups
of the elements $x_\pm$, we identify the orbits ${\cal O}_\pm$ with the
homogeneous manifolds $\tilde H/\tilde H_\pm$. More precisely, we
establish such identification putting into the correspondence to a coset
$h \tilde H_\pm$, $h \in \tilde H$, the element $x(h \tilde H_\pm) \in
{\cal O}_\pm$ given by
\begin{equation}
x(h \tilde H_\pm) = \sqrt{-1} \Ad (h) x_{\pm}.
\end{equation}
The factor $\sqrt{-1}$ is introduced in this relation for the future
convenience.

Let ${\cal O}_{\pm g}$ be a subset of $T_g^{(1,0)}(G)$ defined by
\begin{equation}
{\cal O}_{\pm g} \equiv L_{g*}({\cal O}_\pm).
\end{equation}
We call a mapping $\varphi:M \to G$ an {\it admissible mapping}, if it
satisfies the grading condition, and, moreover, $\varphi_{-*}(\partial_{z
p}) \in \pi_{-*}({\cal O}_{-g})$ and $\varphi_{+*}(\partial_{\bar z p})
\in \pi_{+*}({\cal O}_{+g})$ for any $p \in M$ and $g \in G$ such that
$\pi_\pm (g) = p$. Due to the properties of the orbits ${\cal O}_\pm$
discussed above, this definition does not depend on the choice of the
local coordinate $z$. If the mapping $\varphi$ is admissible, then we can
write
\begin{equation}
{}^{\mu_-} \omega = \lambda_- dz, \qquad {}^{\mu_+} \omega =
\lambda_+ d\bar z, \label{2.96}
\end{equation}
where $\mu_\pm$ are the mappings arising from the local decompositions
(\ref{2.92}), (\ref{2.93}), and $\lambda_\pm$ are the mappings taking
values in $\tilde H/\tilde H_\pm$.  The mappings $\lambda_\pm$ are defined
in the open set $V$ from Corollary \ref{c:2.2}. It follows from
Proposition \ref{p:2.9}, that the mapping $\lambda_-$ is holomorphic, and the
mapping $\lambda_+$ is antiholomorphic. Let $\gamma_\pm$ be local lifts of
the mappings $\lambda_\pm$ to the group $\tilde H$. These mappings are
defined in some open set $W \subset V$, and satisfy the relations
\begin{equation}
\lambda_\pm = \sqrt{-1} \Ad (\gamma_\pm) x_\pm. \label{2.97}
\end{equation}
Note that in the case when the groups $\tilde H_\pm$ are nontrivial, the
mappings $\gamma_\pm$ are defined ambiguously, but in any case they can be
chosen in such a way that the mapping $\lambda_-$ would be holomorphic,
and the mapping $\lambda_+$ would be antiholomorphic.  In what follows we
will use in our consideration such a choice.

\begin{thm} \label{t:2.3}
Let $\varphi: M \to G$ be an admissible mapping.  There exists a local
$\tilde H$--gauge transformation that transforms a connection ${}^\varphi
\omega$ to the connection ${}^{\varphi'}\omega$ of the form
\begin{equation}
{}^{\varphi'} \omega =  (\sqrt{-1} x_- + ({}^\gamma \omega)_-) dz +
\sqrt{-1} \Ad (\gamma^{-1}) x_+ d \mbar z, \label{2.98}
\end{equation}
where $\gamma$ is the mapping taking values in $\tilde H$ and given by
\begin{equation}
\gamma \equiv \gamma_+^{-1}  \eta  \gamma_-
\label{2.99}
\end{equation}
with
\begin{equation}
\eta \equiv \eta_-  \eta_+^{-1}.
\end{equation}
\end{thm}

\begin{pf}
Using representation (\ref{2.93}) and Proposition \ref{p:2.1}, we can write
\begin{equation}
{}^\varphi \omega = {}^{\mu_- \nu_+  \eta_+} \omega
= \Ad (\eta_+^{-1}  \nu_+^{-1})({}^{\mu_-}\omega) +
\Ad (\eta_+^{-1}) ({}^{\nu_+} \omega) + {}^{\eta_+} \omega. \label{2.100}
\end{equation}
In the same way, representation (\ref{2.92}) gives
\begin{equation}
{}^\varphi \omega = {}^{\mu_+ \nu_-  \eta_-} \omega
= \Ad (\eta_-^{-1}  \nu_-^{-1})({}^{\mu_+}\omega) +
\Ad (\eta_-^{-1}) ({}^{\nu_-} \omega) + {}^{\eta_-} \omega; \label{2.101}
\end{equation}
and since the form ${}^{\mu_-} \omega$ is holomorphic, and the form
${}^{\mu_+} \omega$ is antiholomorphic,
\begin{equation}
({}^{\mu_-} \omega)_+ = 0, \qquad ({}^{\mu_+} \omega)_- = 0. \label{2.103}
\end{equation}
Taking into account these relations, we come to the following consequences
of Eqs.~(\ref{2.100}) and (\ref{2.101}):
\begin{equation}
({}^\varphi \omega)_+ = \Ad (\eta_+^{-1})({}^{\nu_+}
\omega)_+ + ({}^{\eta_+} \omega)_+, \qquad
({}^\varphi \omega)_- = \Ad (\eta_-^{-1})({}^{\nu_-}
\omega)_- + ({}^{\eta_-} \omega)_-. \label{2.105}
\end{equation}

Consider now the mapping
\begin{equation}
\kappa \equiv \mu_+^{-1}  \mu_-. \label{2.106}
\end{equation}
Proposition \ref{p:2.1} and Corollary \ref{c:2.1} provide the relation
\begin{equation}
{}^\kappa \omega = {}^{\mu_-} \omega - \Ad (\kappa^{-1}) ({}^{\mu_+}
\omega).  \label{2.107}
\end{equation}
Using Eq.~(\ref{2.96}), we come from this relation to the equalities
\begin{equation}
({}^\kappa \omega)_+ = - \Ad (\kappa^{-1}) \lambda_+, \qquad
({}^\kappa \omega)_- = \lambda_-. \label{2.108}
\end{equation}
Thanks to decompositions (\ref{2.92}) and (\ref{2.93}), we
conclude that the mapping $\kappa$ can be also represented as
\begin{equation}
\kappa = \nu_-  \eta  \nu_+^{-1}. \label{2.109}
\end{equation}
Representation (\ref{2.109}) leads to the equality
\begin{equation}
{}^\kappa \omega = \Ad (\nu_+  \eta^{-1}) ({}^{\nu_-} \omega -
{}^{\eta^{-1}} \omega - \Ad (\eta)({}^{\nu_+} \omega)), \label{2.111}
\end{equation}
which results in the formula
\begin{equation}
\Ad (\eta  \nu_+^{-1}) \lambda_- = ({}^{\nu_-} \omega)_- -
({}^{\eta^{-1}} \omega)_- - \Ad (\eta)({}^{\nu_+} \omega)_-. \label{2.112}
\end{equation}
Taking the ${\goth n}_-$--component of Eq.~(\ref{2.112}), we come to the
relation
\begin{equation}
({}^{\nu_-} \omega)_- = \Ad (\eta) \lambda_-. \label{2.113}
\end{equation}
Similarly, it follows from Eq.~(\ref{2.111}) that
\begin{equation}
- \Ad (\eta^{-1}  \nu_-^{-1}) \lambda_+ = \Ad (\eta^{-1})
({}^{\nu_-} \omega)_+ + ({}^\eta \omega)_+ - ({}^{\nu_+} \omega)_+,
\label{2.114}
\end{equation}
and ${\goth n}_+$--component of this relation is
\begin{equation}
({}^{\nu_+} \omega)_- = \Ad (\eta^{-1}) \lambda_+. \label{2.115}
\end{equation}

Substitution of Eqs.~(\ref{2.113}) and (\ref{2.115}) into
Eq.~(\ref{2.105}) gives
\begin{equation}
({}^\varphi \omega)_+ = \Ad (\eta_-^{-1}) \lambda_+ +
({}^{\eta_+} \omega)_+, \qquad
({}^\varphi \omega)_- = \Ad (\eta_+^{-1}) \lambda_- +
({}^{\eta_-} \omega)_-. \label{2.116}
\end{equation}
Taking into account Eq.~(\ref{2.97}), we come to the relations
\begin{equation}
({}^\varphi \omega)_+ = \sqrt{-1} \Ad (\eta_-^{-1}  \gamma_+) x_+ +
({}^{\eta_+} \omega)_+, \qquad
({}^\varphi \omega)_- = \sqrt{-1} \Ad (\eta_+^{-1}  \gamma_-) x_- +
({}^{\eta_-} \omega)_-. \label{2.117}
\end{equation}
Performing now the gauge transformation defined by the mapping
$\eta_+^{-1}  \gamma_-$, we arrive at representation (\ref{2.98}) with
the mapping $\gamma$ given by Eq.~(\ref{2.99}).
\end{pf}

For the case of a linear group one can rewrite relation (\ref{2.98}) in the
form
\begin{equation}
{}^\varphi \omega =  \left(\sqrt{-1} x_- + \gamma^{-1} \partial_z \gamma
\right) dz + \sqrt{-1} (\gamma^{-1} x_+ \gamma) d\bar z, \label{2.118}
\end{equation}
which allows to present the zero curvature condition (\ref{2.3}) in the
form
\begin{equation}
\partial_{\bar z} \left( \gamma^{-1} \partial_z \gamma \right) =
[\gamma^{-1} x_+ \gamma, x_-].\label{2.119}
\end{equation}
In a general case, instead of Eq.~(\ref{2.119}) we have the following
equation
\begin{equation}
\partial_{\bar z} (({}^\gamma \omega)_-) = [\Ad (\gamma^{-1})
x_+, x_-], \label{2.120}
\end{equation}
which is given here without proof. In concrete examples we usually deal
with linear groups and use Eq.~(\ref{2.119}).

The system of equations for parameters of the group $\tilde H$, which
follows from Eq.~(\ref{2.119}), is called the {\it Toda equations}. In the case
of the canonical gradation, the subgroup $\tilde H$ coincide with some
Cartan subgroup $H$ of $G$, and is, by this reason, an abelian subgroup.
The corresponding equations in this case are called the {\it abelian Toda}
equations. In the case of a general ${\Bbb Z}$--gradation associated with
an integral embedding of $\mbox{\goth sl}(2, {\Bbb C})$ into ${\goth g}$, the
subgroup $\tilde H$ is not necessarily abelian, and we deal with the
abelian Toda equations and their nonabelian versions.

Let us derive the concrete form of the abelian Toda equations.  These
equations will be used in section 4 to prove the generalised Pl\"ucker
relations. In the case under consideration, we can locally parametrise
the mapping $\gamma$ by the set of complex functions $f_i$ as
\begin{equation}
\gamma = \exp \left(\sum_{i=1}^r f_i h_i\right), \label{5.1}
\end{equation}
where $r = \rank {\goth g}$, and the elements $h_i \in {\goth h}$
correspond to the simple roots of ${\goth g}$ (see appendix B).  Choose as the
elements $x_\pm \in {\goth g}_{\pm 1}$ the elements describing the
corresponding principal embedding of $\mbox{\goth sl}(2, {\Bbb C})$ in ${\goth
g}$. The concrete form of such elements is given by Eq.~(\ref{b.39}).
Using Eq.~(\ref{b.9}), we get the relation
\begin{equation}
\gamma^{-1} x_+ \gamma = \sum_{i = 1}^r (2k_i)^{1/2} \exp [- (kf)_i]
x_{+i}, \label{5.2}
\end{equation}
where
\begin{equation}
(kf)_i \equiv  \sum_{j=1}^r k_{ij} f_j. \label{5.3}
\end{equation}
{}From this relation we immediately have
\begin{equation}
[\gamma^{-1} x_+ \gamma, x_-] = \sum_{i=1}^r 2k_i \exp [- (kf)_i] h_i.
\label{5.4}
\end{equation}
{}From the other hand, it is clear that
\begin{equation}
\partial_{\bar z} \left( \gamma^{-1} \partial_z \gamma \right) =
\sum_{i=1}^r (\partial_{\bar z} \partial_z f_i) h_i. \label{5.5}
\end{equation}
Thus, in the case under consideration, equations (\ref{2.119}) can be
reduced to the system
\begin{equation}
\partial_{\bar z} \partial_z f_i = 2k_i \exp [- (kf)_i],
\qquad i = 1, \ldots, r. \label{5.6}
\end{equation}
Introducing the functions
\begin{equation}
u_i \equiv (kf)_i - \ln(2k_i), \label{5.7}
\end{equation}
we rewrite equations (\ref{5.6}) in the form
\begin{equation}
\partial_{\bar z} \partial_z u_i = \sum_{j=1}^r k_{ij} \exp (- u_j),
\label{5.8}
\end{equation}
which is standard for the abelian Toda equations.

\subsection{Gauge Invariance and Dependence on Lifts}

Consider now the behaviour of the mapping $\gamma$ under the $\tilde
H$--gauge transformation. Let $\varphi' = \varphi  \psi$, where the
mapping $\psi$ takes values in $\tilde H$.  It is clear that to define the
mapping $\gamma'$ corresponding to the mapping $\varphi'$, we can use the
same modified Gauss decompositions, which we have used for the construction
of the mapping $\gamma$. From the relations (\ref{2.94}) and (\ref{2.82})
we have $\mu'_+ = \mu_+$. Using the same lift from $\tilde H/\tilde H_+$
to $\tilde H$, which was used in transition from the mapping $\mu_+$ to the
mapping $\gamma_+$, we obtain $\gamma'_+ = \gamma_+$.  From the other
hand, relations (\ref{2.95}) and (\ref{2.84}) give $\eta'_- = \eta_-
\psi$. In the similar way we have $\gamma'_- = \gamma_-$, and $\eta'_+ =
\eta_+ \psi$. Thus, it follows from Eq.~(\ref{2.99}) that $\gamma' =
\gamma$. Just in this sense the mapping $\gamma$ is $\tilde H$--gauge
invariant.

The last question we are going to consider in this section, is the
dependence of the mapping $\gamma$ on the choice of modified Gauss
decompositions and local lifts from $\tilde H/\tilde H_\pm$ to $\tilde H$.
Let we have two local decompositions of the mapping $\varphi$:
\begin{equation}
\varphi = \mu_+  \nu_-  \eta_-, \qquad \varphi = \mu'_+
\nu'_-  \eta'_-, \label{2.121}
\end{equation}
which are obtained with the help of the modified Gauss decompositions
corresponding to the indexes $a$ and $b$, respectively. Using
Eqs.~(\ref{2.79b}) and (\ref{2.81}), we get
\begin{equation}
\mu'_+ = m_{+b} \circ \varphi = \mu_+  \nu_{-ab}  \eta_{-ab},
\label{2.122}
\end{equation}
where
\begin{equation}
\nu_{-ab} \equiv n_{-ab} \circ \varphi_+, \qquad \eta_{-ab} \equiv h_{-ab}
\circ \varphi_+. \label{2.123}
\end{equation}
It is evident that the mappings $\nu_{-ab}$ and $\eta_{-ab}$ are
holomorphic, and take values in $\tilde N_-$ and $\tilde H$, respectively.
{}From Eq.~(\ref{2.122}) it follows that
\begin{equation}
{}^{\mu'_+} \omega = \Ad (\eta_{-ab}^{-1})(
\Ad (\nu_{-ab}^{-1})({}^{\mu_+} \omega) + {}^{\nu_{-ab}} \omega) +
{}^{\eta_{-ab}} \omega. \label{2.124}
\end{equation}
Since the forms ${}^{\mu_+} \omega$ and ${}^{\mu'_+} \omega$ take values
in ${\goth g}'_{+1}$, we actually have the relation
\begin{equation}
{}^{\mu'_+} \omega = \Ad (\eta_{-ab}^{-1})({}^{\mu_+} \omega).
\label{2.125}
\end{equation}
Taking into account Eqs.~(\ref{2.96}) and (\ref{2.97}), we get
\begin{equation}
\Ad (\gamma'_+) x_+ = \Ad (\eta_{-ab}  \gamma_+) x_+.
\label{2.126}
\end{equation}
It follows from this relation that
\begin{equation}
\gamma'_+ = \eta_{-ab}^{-1}  \gamma_+  \xi_+, \label{2.127}
\end{equation}
where the mapping $\xi_+$ takes values in $\tilde H_+$. It is clear that the
mapping $\xi_+$ is antiholomorphic.

Further, Eq.~(\ref{2.84b}) allows to write
\begin{equation}
\eta_- = \eta_{-ab}  \eta'_-. \label{2.128}
\end{equation}
Combining now Eqs.~(\ref{2.127}) and (\ref{2.128}), we obtain
\begin{equation}
\gamma^{\prime -1}_+  \eta'_- = \xi^{-1}_+  \gamma^{-1}_+
\eta_-.
\end{equation}
In the similar way we have
\begin{equation}
\eta^{\prime -1}_+  \gamma'_- = \eta^{-1}_+  \gamma_-  \xi_-,
\end{equation}
where $\xi_-$ is a holomorphic mapping taking values in $\tilde H_-$.
Finally, we come to the relation
\begin{equation}
\gamma' = \xi_+^{-1}  \gamma  \xi_-.
\end{equation}

Resuming our consideration, we can say that any admissible mapping
$\varphi$ leads to a set $\{\gamma_i\}_{i \in I}$ of local solutions of
equations (\ref{2.119}).  These solutions, in the overlaps of their domains,
are connected by the relations
\begin{equation}
\gamma_i = \xi^{-1}_{+ij}  \gamma_j  \xi_{-ij},
\end{equation}
where the mappings $\xi_{-ij}$ are holomorphic and take values in $\tilde
H_-$, while the mappings $\xi_{+ij}$ are antiholomorphic and take values
in $\tilde H_+$.  Note here that Eq.~(\ref{2.127}) describes symmetry
transformations of equations (\ref{2.119}). For the case of the canonical
gradation, the subgroups $\tilde H_\pm$ are discrete and isomorphic to
$\underbrace{{\Bbb Z}_2 \times \cdots \times {\Bbb Z}_2}_r$.

\section{Construction of Solutions and Reality Condition}

In the previous section we have shown that any admissible mapping $\varphi: M
\to G$ allows to construct a set of local solutions of equations
(\ref{2.119}). From the first glance, the problem of constructing admissible
mappings is rather complicated. In fact, any admissible mapping $\varphi$
is defined by the corresponding mappings $\varphi_+$ and $\varphi_-$ up to
an $\tilde H$--gauge transformation. In other words the following
proposition is valid.

\begin{pro} \label{p:3.1}
Let $\varphi$ and $\varphi'$ be two mappings from $M$ to $G$. The equalities
\begin{equation}
\varphi'_+ = \varphi_+, \qquad \varphi'_- = \varphi_- \label{3.1}
\end{equation}
are valid if and only if $\varphi' = \varphi \psi$, where the
mapping $\psi$ takes values in $\tilde H$.
\end{pro}

\begin{pf}
The assertion of the propositions follows directly from the fact that
$\tilde B_- \cup \tilde B_+ = \tilde H$.
\end{pf}

Since the mapping $\gamma$ is $\tilde H$--gauge invariant, we conclude
that this mapping can be constructed using only information contained in
the mappings $\varphi_\pm$. In practice, it can be done in the following
way.

Let $\sigma$ be an involutive antilinear automorphism of ${\goth g}$.
It is known that the mapping $\sigma$ defines a real form
${\goth g}_\sigma$ of the Lie algebra ${\goth g}$ by
\begin{equation}
{\goth g}_\sigma \equiv \{ x \in {\goth g} \mid \sigma(x) = x\}.
\label{3.2}
\end{equation}
Introduce the $\sigma$--hermitian conjugation as a mapping from ${\goth
g}$ to ${\goth g}$ which assigns to any element $x \in {\goth g}$ the
element $x^\star \in {\goth g}$, defined by
\begin{equation}
x^\star \equiv - \sigma(x). \label{3.3}
\end{equation}
It is clear that the $\sigma$--hermitian conjugation is an antilinear
antiautomorphism of ${\goth g}$ satisfying the condition
\begin{equation}
(x^\star)^\star = x \label{3.4}
\end{equation}
for any $x \in {\goth g}$. An element $x \in {\goth g}$ is said to be
$\sigma$--hermitian if $x^\star = x$, and it is said to be
$\sigma$--antihermitian if $x^\star = -x$. The subalgebra
${\goth g}_\sigma$ is formed by all $\sigma$--antihermitian elements of
${\goth g}$.

Suppose that $\sigma$ can be extended to an antiholomorphic automorphism
$\Sigma$ of the group $G$. In this case we have $\Sigma^2 = \id_G$.
The $\Sigma$--hermitian conjugation in $G$ is defined then as a mapping $g
\in G \to g^\star \in G$, where
\begin{equation}
g^\star \equiv \Sigma(g^{-1}) = (\Sigma(g))^{-1}. \label{3.5}
\end{equation}
It is evident that $\Sigma$--hermitian conjugation is an antiholomorphic
antiautomorphism of $G$ satisfying the condition
\begin{equation}
(g^\star)^\star = g \label{3.6}
\end{equation}
for any $g \in G$. An element $g \in G$ is called {\it $\Sigma$--hermitian} if
$g^\star = g$, and it is called {\it $\Sigma$--unitary} if $g^\star = g^{-1}$.
In the case when the mapping $\Sigma$ is originated from the mapping
$\sigma$, giving a compact real form of ${\goth g}$, we say simply
`hermitean' and `unitary'.  The real Lie group $G_\sigma$ corresponding
to the real form ${\goth g}_\sigma$, is formed by all $\Sigma$--unitary
elements of $G$.

Suppose also that the mapping $\sigma$ has the property
\begin{equation}
\sigma({\goth g}_m) = {\goth g}_{-m}, \qquad m \in {\Bbb Z}. \label{3.7}
\end{equation}
In this case one has
\begin{equation}
\tilde H^\star = \tilde H, \qquad (\tilde N_{\pm})^\star = \tilde
N_{\mp}. \label{3.8}
\end{equation}

A representation $\rho: G \to GL(V)$ of the group $G$ in the linear
space $V$ over the field ${\Bbb C}$ is called {\it $\Sigma$--unitary}, if the
space $V$ is equipped with a hermitean scalar product $\langle\ ,\
\rangle$ such that
\begin{equation}
\rho(g)^\star = \rho(g^\star), \label{3.9}
\end{equation}
where $\star$ in the left hand side means the hermitean conjugation with
respect to the scalar product $\langle\ , \ \rangle$. It can be shown that
the representation $\rho$ is $\Sigma$--hermitean if and only if the
restriction of $\rho$ to the real Lie group $G_\sigma$ is unitary.

Let us consider now an arbitrary $\Sigma$--unitary representation of the
group $G$ in a linear space $V$. Denote by $V_+$ the subspace of $V$
formed by all elements $v\in V$, such that
\begin{equation}
\rho(g)\,v = v \label{3.10}
\end{equation}
for all $g \in \tilde N_+$. For any mapping $\varphi: M \to G$ and any vectors
$u,\, v \in V$ we denote by $\langle u,\, \varphi\, v\rangle$ the mapping
from $M$ to ${\Bbb C}$, defined as
\begin{equation}
\langle u,\, \varphi\, v \rangle (p) \equiv \langle u,\, \rho(\varphi(p))\,
v\rangle. \label{3.11}
\end{equation}

\begin{thm} \label{t:3.1}
For any $u, v \in V_+$ the following relation is valid:
\begin{equation}
\langle u,\, \gamma\, v \rangle = \langle u,\, (\mu_+ \gamma_+)^{-1}
(\mu_- \gamma_-)\, v \rangle. \label{3.12}
\end{equation}
\end{thm}

\begin{pf}
Comparing Eqs.~(\ref{2.106}) and (\ref{2.109}), we get
\begin{equation}
\nu_- \eta \nu_+^{-1} = \mu_+^{-1} \mu_-. \label{3.13}
\end{equation}
Now, using the definition (\ref{2.99}) of the mapping $\gamma$, we come to
the relation
\begin{equation}
\nu'_- \gamma \nu_+^{\prime -1} = (\mu_+ \gamma_+)^{-1} (\mu_- \gamma_-),
\label{3.14}
\end{equation}
where
\begin{equation}
\nu'_- \equiv \gamma_+^{-1} \nu_- \gamma_+, \qquad \nu'_+ \equiv
\gamma_-^{-1} \nu_+ \gamma_-. \label{3.15}
\end{equation}
Then, the validity of Eq.~(\ref{3.12}) follows from Eq.~(\ref{3.14}) and
the definition of $V_+$.
\end{pf}

Thus, we can find some matrix elements of the linear operators
corresponding to the mapping $\gamma$ in a $\Sigma$--unitary
representation of the group $G$, using only the mappings $\mu_\pm$, which
in their turn are determined only by the mappings $\varphi_\pm$. It is
natural to suppose here that using a reach enough set of representations,
one will be able to recover the mapping $\gamma$ from the mappings
$\varphi_\pm$. It is likely that for the case of simple Lie groups, it is
enough to use only fundamental representations, that is confirmed by the
consideration of concrete examples.

Let us consider now a special class of the solutions which will be called
{\it real solutions}. To this end, introduce two antiholomorphic mappings
$\Sigma_\pm : F_\pm \to F_\mp$ defined by
\begin{equation}
\Sigma_\pm(g \tilde B_\mp) \equiv \Sigma(g) \tilde B_\pm \label{3.16}.
\end{equation}
It is easy to get convinced that the mappings $\Sigma_\pm$ are
defined correctly. Directly from the definition of these mappings, we get
the equalities
\begin{equation}
\Sigma_+ \circ \pi_+ = \pi_- \circ \Sigma, \qquad \Sigma_- \circ \pi_- =
\pi_+ \circ \Sigma. \label{3.17}
\end{equation}
Moreover, these mappings are mutually inverse:
\begin{equation}
\Sigma_+ \circ \Sigma_- = \id_{F_-}, \qquad \Sigma_- \circ \Sigma_+ =
\id_{F_+}. \label{3.18}
\end{equation}
A mapping $\varphi: M \to G$ is said to satisfy the reality condition if
\begin{equation}
\Sigma_+ \circ \varphi_+ = \varphi_-, \label{3.19}
\end{equation}
that can be also written as
\begin{equation}
\Sigma_- \circ \varphi_- = \varphi_+. \label{3.20}
\end{equation}

\begin{pro} \label{p:3.2}
A mapping $\varphi: M \to G$ satisfies the reality condition if and only
if
\begin{equation}
\Sigma \circ \varphi = \varphi \psi, \label{3.21}
\end{equation}
where the mapping $\psi$ takes values in $\tilde H$.
\end{pro}

\begin{pf}
Let $\varphi$ satisfies the reality condition. Using
Eq.~(5.17) in Eq.~(\ref{3.19}), we get
\begin{equation}
\pi_- \circ \Sigma \circ \varphi = \pi_- \circ \varphi. \label{3.22}
\end{equation}
Hence, $\Sigma \circ \varphi = \varphi\psi$, where $\psi$ takes values in
$\tilde B_+$. In the similar way, from Eq.~(\ref{3.20}) we see that
$\psi$ takes values in $\tilde B_-$. Since $\tilde B_+ \cap
\tilde B_- = \tilde H$, we conclude that $\psi$ takes values in
$\tilde H$.

The inverse statement of the proposition is evident.
\end{pf}

A mapping $\psi: M \to G$ is called {\it $\Sigma$--hermitean} if for any
$p \in M$ the mapping $\psi(p)$ is a $\Sigma$--hermitean element of $G$.

\begin{pro} \label{p:3.3}
The mapping $\psi$ from proposition {\rm \ref{p:3.2}} is $\Sigma$--hermitean.
\end{pro}

\begin{pf}
Since $\Sigma^2 = \id_G$, from Eq.~(\ref{3.21}) we get
\begin{equation}
\varphi = (\Sigma \circ \varphi)(\Sigma \circ \psi), \label{3.23}
\end{equation}
that can be written as
\begin{equation}
\Sigma \circ \varphi = \varphi (\Sigma \circ \psi)^{-1}. \label{3.24}
\end{equation}
Comparing Eq.~(\ref{3.24}) with Eq.~(\ref{3.21}) we have
\begin{equation}
(\Sigma \circ \psi)^{-1} = \psi. \label{3.25}
\end{equation}
Hence, the mapping $\psi$ is $\Sigma$--hermitean.
\end{pf}

\begin{pro} \label{p:3.4}
The reality condition is $\tilde H$--gauge invariant.
\end{pro}

\begin{pf}
Let a mapping $\varphi$ satisfies the reality condition, and $\varphi' =
\varphi \xi$ with $\xi$ taking values in $\tilde H$, be a gauge
transformed mapping. Using Eq.~(\ref{3.21}) we have
\begin{equation}
\Sigma \circ \varphi' = \varphi' \circ \psi', \label{3.26}
\end{equation}
where
\begin{equation}
\psi' = \xi^{-1} \psi (\Sigma \circ \xi). \label{3.27}
\end{equation}
Since $\Sigma (\tilde H) = \tilde H$, the mapping $\psi'$
satisfies the reality condition.
\end{pf}

Let $s_{+a}$, $a \in A$, be a family of local holomorphic sections of the
fibre bundle $G \to F_+$ with the properties described in proposition 4.3.
The mapping $\Sigma$ allows us to construct the corresponding family of
local holomorphic sections of the fibre bundle $G \to F_-$. For
each open set $U_{+a}$ we define the open set $U_{-a}$ by
\begin{equation}
U_{-a} \equiv \Sigma_+(U_{+a}). \label{3.28}
\end{equation}
Using Eqs.~(\ref{3.17}) and (\ref{3.18}), it is easy to get convinced that
for any $a\in A$, the mapping $s_{_a}: U_{-a} \to G$ given by
\begin{equation}
s_{-a} \equiv \Sigma \circ s_{+a} \circ \Sigma_-, \label{3.29}
\end{equation}
is a local holomorphic section of the fibre bundle $G \to F_-$. Since
$\Sigma(\tilde N_+) = \tilde N_-$, we also have
\begin{equation}
\Sigma_*{\cal N}_+ = {\cal N}_-; \label{3.30}
\end{equation}
hence, the set $s_{-a}(U_{-a})$ is an integral manifold of the
distribution ${\cal N}_-$. Thus, we get a family of holomorphic
sections of the fibre bundle $G \to F_-$ with the required properties.
Now, to construct the mappings needed to define the mapping $\gamma$, we
will use for any section $s_{+a}$ the corresponding section $s_{-a}$
defined with the help of the procedure described above.

\begin{pro} \label{p:3.5}
If a mapping $\varphi$ satisfies the reality condition, then the mappings
$\mu_+$ and $\mu_-$ entering proposition {\rm \ref{p:2.8}} are connected by
\begin{equation}
\mu_\mp = \Sigma \circ \mu_\pm. \label{3.31}
\end{equation}
\end{pro}

\begin{pf}
Recall that the mappings $\mu_\pm$ are given by
\begin{equation}
\mu_\pm = s_{\pm a} \circ \varphi_\pm. \label{3.32}
\end{equation}
{}From this relation we have
\begin{equation}
\mu_\mp = \Sigma \circ s_{\pm a} \circ \Sigma_\mp \circ \varphi_\mp =
\Sigma \circ s_{\pm a} \circ \varphi_\pm = \Sigma \circ \mu_\pm.
\label{3.33}
\end{equation}
This chain of equalities provides the assertion of the proposition.
\end{pf}

\begin{pro} \label{p:3.6}
If a mapping $\varphi$ satisfies the reality condition, then the mappings
$\gamma_\pm$ satisfying Eq.~{\rm (\ref{2.97})} can be chosen in such a way
that
\begin{equation}
\Sigma \circ \gamma_\pm = \gamma_\mp. \label{3.34}
\end{equation}
\end{pro}

\begin{pf}
First, let us show that the mappings $\lambda_+$ and $\lambda_-$ are
connected by
\begin{equation}
\lambda_- = \sigma \circ \lambda_+. \label{3.35}
\end{equation}
{}From Eq.~(\ref{3.31}) we have
\begin{equation}
{}^{\mu_-} \omega = (\mu_+^* \circ \Sigma^*) \theta. \label{3.36}
\end{equation}
Now, using the equality (\ref{a.30}), we get
\begin{equation}
{}^{\mu_-}\omega(x) = \Sigma^* \theta(\mu_{+*}(x)) = \sigma\left ({}^{\mu_+}
\omega (\mbar x)\right) \label{3.38}
\end{equation}
for any $x \in T^{\Bbb C}_p(M)$, $p \in M$. In particular, we can write
the equality
\begin{equation}
{}^{\mu_-}\omega (\partial_z) = \sigma \left({}^{\mu_+} \omega
(\partial_{\bar z})\right), \label{3.39}
\end{equation}
which directly leads to Eq.~(\ref{3.35}).

{}From Eqs.~(\ref{3.35}) it follows that
\begin{equation}
\Ad (\gamma_-) x_- = (\sigma \circ \Ad (\gamma_+)) x_+.
\label{3.40}
\end{equation}
Suppose now that the elements $x_\pm$ entering Eq.~(\ref{2.97}) are chosen
in such a way that
\begin{equation}
\sigma(x_+) = -x_-. \label{3.41}
\end{equation}
Taking into account relation (\ref{a.6b}), we conclude that one can choose
the mappings $\gamma_+$ and $\gamma_-$ satisfying Eq.~(\ref{3.34}).
\end{pf}

\begin{thm} \label{t:3.2}
If a mapping $\varphi$ satisfies the reality condition, then the mapping
$\gamma$ can be chosen to be $\Sigma$--hermitean.
\end{thm}

\begin{pf}
{}From proposition \ref{p:3.2}, using Eqs.~(\ref{2.92}) and (\ref{2.93}), we
get
\begin{equation}
\Sigma \circ \varphi = (\Sigma \circ \mu_+)(\Sigma \circ \nu_-)(\Sigma
\circ \eta_-) = \mu_- \nu_+ \eta_+ \psi. \label{3.43}
\end{equation}
Since decomposition (\ref{2.93}) is unique, we have
\begin{equation}
\Sigma \circ \eta_- = \eta_+ \psi, \label{3.44}
\end{equation}
and hence, can write the equality
\begin{equation}
\eta_- = (\eta_+^{-1})^\star \psi^{-1}, \label{3.45}
\end{equation}
which leads to the relation
\begin{equation}
\eta = (\eta_+^{-1})^\star \psi^{-1} \eta_+^{-1}. \label{3.46}
\end{equation}
As it follows from proposition \ref{p:3.3}, the mapping $\psi$ is
$\Sigma$--hermitean, thus the mapping $\eta$ is also $\Sigma$--hermitean.
Taking into account  proposition \ref{p:3.6} and the definition of the
mapping $\gamma$, we conclude that it can be chosen to be
$\Sigma$--hermitean.
\end{pf}

Thus, we can say that using the mappings $\varphi$ satisfying the reality
condition, one can construct hermitean, in a sense real, solutions of the
equations  (\ref{2.119}). Since the solutions are actually determined by
the mappings $\varphi_\pm$, then to get real solutions of the equations
under consideration, we should choose the mappings $\varphi_\pm$
satisfying relation (\ref{3.19}). Note that, in general, not any
pair of the mappings $\varphi_\pm$ can be used for the construction of a
solution. Actually, there are mappings $\varphi_\pm$ which do not
correspond to any mapping $\varphi$. We call mappings
$\varphi_\pm: M \to F_\pm$ {\it consistent} if there is a mapping $\varphi: M
\to G$ such that $\varphi_\pm = \pi_\pm \circ \varphi$.

Suppose now that the involutive antilinear automorphism $\sigma$ is
defined by Eq.~(\ref{b.7}), so that the corresponding real form ${\goth
u}$ of ${\goth g}$ is compact. In this case the gradation operator is
$\sigma$--hermitean, and Eq.~(\ref{3.7}) is valid.

\begin{thm} \label{t:3.3}
If mappings $\varphi_\pm: M \to F_\pm$ satisfy relation {\rm
(\ref{3.19})}, then they are consistent.
\end{thm}

\begin{pf}
It is enough to show that if two points $p_+ \in F_+$ and $p_- \in
F_-$ are connected by the relation
\begin{equation}
p_- = \Sigma_+(p_+);
\end{equation}
then there exists an element $g \in G$ such that
\begin{equation}
p_+ = \pi_+(g), \qquad p_- = \pi_-(g).
\end{equation}
This fact can be proved using the Iwasawa decomposition (\ref{b.15}). Let
$g'$ be any element of $G$, such that $\pi_+(g') = p_+$. This element can
be written as
\begin{equation}
g' = u n_+ a,
\end{equation}
where $u \in U$, $n_+ \in N_+$ and $a \in A^*$, and the subgroups $U$,
$N_+$ and $A^*$ are defined in appendix B. We have $N_+ A \subset B_+
\subset \tilde B_+$, hence $\pi_+ (u) = p_+$. Using now
Eq.~(\ref{3.17}) one gets
\begin{equation}
(\Sigma_+ \circ \pi_+)(u) = (\pi_- \circ \Sigma)(u) = \pi_-(u) =
\Sigma_+(p_+).
\end{equation}
Thus, the element $u$ can be taken as the element $g$ we are looking for.
\end{pf}

The generalisation of the results proved in this section and in the
previous one, to the case of the semi--integral embeddings of ${\goth
sl}(2, {\Bbb C})$ into ${\goth g}$ is straightforward, and can be
performed following the consideration given in Ref.~\cite{LeS92}. Recall
that in that case we deal with ${\Bbb Z}/2$--gradations of ${\goth g}$
(see appendix~B).

\section{Toda Fields and Generalised Pl\"ucker Relations}

In this section we give a derivation of the generalised Pl\"ucker
relations.  Recall first some definitions. Let $M$ be a complex manifold,
and $J^M$ be the corresponding complex structure on $M$. A metric $g$ on
$M$ is called hermitean if
\begin{equation}
g(J^M X, J^M Y) = g(X, Y) \label{5.9}
\end{equation}
for all vector fields $X$, $Y$ on $M$. Let $g$ be a hermitean metric on
$M$. The 2--form $\Phi$ defined by
\begin{equation}
\Phi(X, Y) \equiv g(X, J^M Y) \label{5.10}
\end{equation}
for all vector fields $X$, $Y$ on $M$ is called the fundamental form
associated with $g$. A hermitean metric $g$ on $M$ is called a K\"ahler
metric if the fundamental form $\Phi$ associated with $g$ is closed, i.e.,
\begin{equation}
d \Phi = 0. \label{5.11}
\end{equation}
The fundamental form $\Phi$ associated with a hermitean metric $g$ can be
locally represented as
\begin{equation}
\Phi = -2\sqrt{-1} \partial \mbar \partial K, \label{5.12}
\end{equation}
where $K$ is a real--valued function. The function $K$ entering
Eq.~(\ref{5.12}), is called a K\"ahler potential of the hermitean metric $g$.

Let $z^i$ be local coordinates in $M$; introduce the notation
\begin{equation}
\partial_{z^i} \equiv \partial/\partial z^i, \qquad \partial_{\bar z^i}
\equiv \partial/\partial \mbar z^i.
\end{equation}
For any hermitean metric $g$ on $M$ we have
\begin{equation}
g(\partial_{z^i}, \partial_{\bar z^j}) = \mbar{g(\partial_{z^j},
\partial_{\bar z^i})}, \label{5.12a}
\end{equation}
while
\begin{equation}
g(\partial_{z^i}, \partial_{z^j}) = 0, \qquad
g(\partial_{\bar z^i}, \partial_{\bar z^j}) = 0.
\label{5.12b}
\end{equation}
Hence, we can write
\begin{equation}
g = \sum_{i, j} g_{i \bar j}(dz^i \otimes d\mbar z^j + d\mbar z^j \otimes
d z_i) \equiv 2 \sum_{i, j} g_{i \bar j} dz^i d\mbar z^j, \label{5.12c}
\end{equation}
where
\begin{equation}
g_{i \bar j} \equiv g(\partial_{z^i}, \partial_{\bar z^j}).
\label{5.12d}
\end{equation}
For the fundamental form $\Phi$ associated with $g$, we get the
representation
\begin{equation}
\Phi = -2 \sqrt{-1} \sum_{i, j} g_{i \bar j} dz^i \wedge d\mbar z^j.
\label{5.12e}
\end{equation}
Note also that the expression for the Ricci curvature tensor $R$ in the case
under consideration has the form \cite{KoN63}
\begin{equation}
R = 2\sqrt{-1} \partial \mbar \partial \ln {\cal G} , \label{5.12f}
\end{equation}
where ${\cal G}$ is the determinant of the matrix $(g_{i \bar j})$.

Let $V$ be a complex linear space, and ${\Bbb P}(V)$ be the projective
space associated with $V$.  In other words, ${\Bbb P}(V)$ is the set of
all lines (one--dimensional subspaces) in $V$. Denote by Pr the canonical
projection from the set $V' \equiv V - \{0\}$ onto ${\Bbb P}(V)$.
Suppose that $V$ is endowed with a hermitean scalar product $\langle\ ,\
\rangle$. Let $F$
be the function on $V'$ given by
\begin{equation}
F (v) \equiv \ln \Vert v \Vert^2, \label{5.13}
\end{equation}
where $\Vert v \Vert^2 \equiv \langle v, v \rangle$. The Fubini--Study
metric on ${\Bbb P}(V)$ is defined by its associated fundamental form
$\Phi^{FS}$, which is a unique form satisfying the relation
\begin{equation}
\Pr^* \Phi^{FS} = -2\sqrt{-1}\partial \mbar \partial F. \label{5.14}
\end{equation}

Let $G$ be a complex semisimple Lie group of rank $r$, ${\goth g}$
be its Lie algebra endowed with the canonical ${\Bbb Z}$--gradation, and
${\goth h}$ be the corresponding Cartan subalgebra. Consider the Borel
subgroup $B_+$, corresponding to the Borel subalgebra ${\goth b}_+$
defined by Eq.~(\ref{b.12}). Let $\varphi_-$ be a holomorphic mapping from
complex manifold $M$ of complex dimension one to the flag manifold $F_- =
G/B_+$. Using the language of the algebraic geometry, we call $\varphi_-$ a
holomorphic curve in $F_-$. For any parabolic subgroup $P$ containing
$B_+$, there is a natural projection from $F_-$ onto the flag
manifold $F_P\equiv G/P$; denote this projection $\pi_P$. Using such a
projection, we can define a holomorphic curve $\varphi_{-P}$ in $F_P$
as
\begin{equation}
\varphi_{-P} \equiv \pi_{-P} \circ \varphi_-. \label{5.15}
\end{equation}
Thus, any holomorphic curve $\varphi_-$ in $F_-$ generates a family of
holomorphic curves $\varphi_{-P}$ labelled by parabolic subgroups
containing the Borel subgroup $B_+$. We are especially interested in
holomorphic curves generated by the parabolic subgroups of the form
$P_{+i'}$, $i = 1,\ldots, r$ (see appendix B).  Denote the corresponding
flag manifolds by $F_{-i}$, and the projections from $F_-$ onto $F_{-i}$
by $\pi_{-i}$. Note that the projections $\pi_{-i}$ are defined by the
relation
\begin{equation}
\pi_{-i}(g B_+) \equiv g P_{+i'}. \label{5.15a}
\end{equation}

The flag manifolds $F_{-i}$ are closely related to the fundamental
representations of the group $G$ and the corresponding representations of
the Lie algebra ${\goth g}$. It is well--known that for a complex
semisimple Lie group of rank $r$ there are $r$ fundamental
representations. Let $\rho_i : G \to \mbox{\rm GL}(V_i)$, $i = 1, \ldots,
r$, be a family of fundamental representations of $G$. We use the notation
$\rho_i$ also for the corresponding fundamental representation of the Lie
algebra ${\goth g}$. Recall that the $i$--th fundamental representation of
the Lie algebra ${\goth g}$ is characterised  by the highest weight vector
$v_i \in V_i$, satisfying the relations
\begin{equation}
\rho_i(h_j) v_i = \delta_{ij}, \qquad j = 1, \ldots, r. \label{5.16}
\end{equation}
Since $v_i$ is the highest weight vector, we have
\begin{equation}
\rho_i(x_{+j}) v_i = 0, \qquad j = 1, \ldots, r. \label{5.17}
\end{equation}
Moreover, it can be shown that
\begin{equation}
\rho_i(x_{-j}) v_i = 0, \qquad i \ne j, \label{5.18}
\end{equation}
and
\begin{equation}
\rho_i(x_{-i}) v_i \ne 0. \label{5.19}
\end{equation}
It follows from Eqs.~(\ref{5.16})--(\ref{5.19}) that the line in $V_i$
containing the vector $v_i$ is invariant under the action of the parabolic
subgroup $P_{+i'}$. Note that, as any representation, the fundamental
representation $\rho_i$ defines a left action of the group $G$ in the
projective space ${\Bbb P}(V_i)$. Here for any $g \in G$ we have
\begin{equation}
g \cdot \Pr_i(v) = \Pr_i (\rho_i(g)v), \qquad v \in V'_i,
\label{5.20}
\end{equation}
where $\Pr_i$ is the natural projection from $V'_i$ onto ${\Bbb
P}(V_i)$. Thus, we can say that the parabolic subgroup $P_{+i'}$ is the
stability subgroup of the point $p_i \in {\Bbb P}(V_i)$, corresponding to
the highest weight vector $v_i$. The above consideration shows that we can
define an embedding $\iota_{F_{-i}}$ of the flag manifold $F_{-i}$
into the projective space ${\Bbb P}(V_i)$ by
\begin{equation}
\iota_{F_{-i}}(g P_{+i'}) \equiv g \cdot p_i. \label{5.21}
\end{equation}
The Fubini--Study metric on ${\Bbb P}(V_i)$ induces a K\"ahler metric on
$F_{-i}$; hence the flag manifold $F_{-i}$ is a K\"ahler manifold. The
K\"ahler metric on $F_{-i}$, via the mapping
\begin{equation}
\varphi_{-i} \equiv \pi_{-i} \circ \varphi_-, \label{5.22}
\end{equation}
specifies a pseudo--metric on $M$. The fundamental form $\Phi_i$ associated
with this pseudo--metric, is given by
\begin{equation}
\Phi_i = (\varphi_-^* \circ \pi_{-i}^* \circ \iota_{F_{-i}}^*)\Phi_i^{FS},
\label{5.23}
\end{equation}
where $\Phi_i^{FS}$ is the fundamental form associated with the
Fubini--Study metric on ${\Bbb P}(V_i)$.

\begin{pro} \label{p:5.1}
Any holomorphic local lift $\tilde \varphi_-$ of the mapping $\varphi_-$ to
$G$ leads to the following local representation of the fundamental form
$\Phi_i$:
\begin{equation}
\Phi_i = -2\sqrt{-1} \partial \mbar \partial \ln \Vert \tilde
\varphi_- v_i \Vert^2.  \label{5.24}
\end{equation}
\end{pro}

\begin{pf}
By definition, the mapping $\tilde \varphi_-$ satisfies the relation
\begin{equation}
\varphi_- = \pi_- \circ \tilde \varphi_-. \label{5.25}
\end{equation}
Define the mapping $\tau_i: G \to V'_i$ by
\begin{equation}
\tau_i(g) \equiv \rho_i(g) v_i, \qquad g \in G, \label{5.26}
\end{equation}
and prove the following equality
\begin{equation}
\Pr_i \circ \tau_i = \iota_{F_{-i}} \circ \pi_{-i} \circ \pi_-.
\label{5.27}
\end{equation}
Indeed, using Eq.~(\ref{5.20}), for any $g \in G$ we get
\begin{equation}
(\Pr_i \circ \tau_i)(g) = \Pr_i(\rho_i(g) v_i) = g \cdot p_i.
\label{5.28}
\end{equation}
{}From the other hand, it follows from Eqs.~(\ref{5.15a}) and (\ref{5.21})
that
\begin{equation}
(\iota_{F_{-i}} \circ \pi_{-i} \circ \pi_-)(g) = \iota_{F_{-i}}(g P_{+i'})
= g \cdot p_i. \label{5.29}
\end{equation}
Hence, equality (\ref{5.27}) is true.

So, Eqs.~(\ref{5.25}) and (\ref{5.26}) results in
\begin{equation}
\iota_{F_{-i}} \circ \pi_{-i} \circ \varphi_- = \Pr_i \circ \tau_i
\circ \tilde \varphi_-. \label{5.30}
\end{equation}
Taking into account Eqs.~(\ref{5.23}) and (\ref{5.14}), we get
\begin{equation}
\Phi_i = (\tilde \varphi_-^* \circ \tau_i^* \circ \Pr_i^*)
\Phi_i^{FS} = -4 \sqrt{-1} \partial \mbar \partial (\tilde \varphi_-^*
\circ \tau_i \circ F).
\label{5.31}
\end{equation}
The assertion of the proposition is the direct consequence of
Eq.~(\ref{5.31}).
\end{pf}

Proposition \ref{p:5.1} shows that the function
\begin{equation}
K_i \equiv \ln \Vert \tilde \varphi_- v_i \Vert^2 \label{5.32}
\end{equation}
is a K\"ahler potential of the pseudo--metric $g_i$ on $M$, having $\Phi_i$
as its fundamental form. Choosing different lifts $\tilde \varphi_-$,
we get different K\"ahler potentials.

Suppose now that the hermitian scalar product in $V_i$ is chosen in such a
way that the corresponding representation of the group $G$ is unitary.
Let $\varphi_{-*}(\partial_{z p}) \in \pi_{-*}({\cal O}_{-g})$ for any $p
\in M$ and $\pi_-(g) = p$.  Consider an arbitrary local lift $\varphi$ of
$\varphi_-$ to $G$, and construct for it a local decomposition of the form
(\ref{2.93}). With the help of Eq.~(\ref{2.97}), define the mapping
$\gamma_-$, corresponding to the mapping $\mu_-$, and the mapping $\tilde
\varphi_-$ by
\begin{equation}
\tilde \varphi_- \equiv \mu_- \gamma_-. \label{5.33}
\end{equation}
It is clear that $\tilde \varphi_-$ is a holomorphic local lift of
$\varphi_-$ to $G$.
It follows from theorem \ref{t:3.1} that for such choice of $\tilde
\varphi_-$ the functions $f_i \equiv K_i$ are real solutions of the Toda
equations (\ref{5.8}).

\begin{thm}
Under the conditions described above, the Ricci curvature tensors $R_i$ of
the pseudo--metrics $g_i$ on $M$ are connected with the corresponding
fundamental forms $\Phi_i$ by the relations
\begin{equation}
R_i = \sum_{j=1}^r k_{ij} \Phi_j. \label{5.34}
\end{equation}
\end{thm}

\begin{pf}
{}From Eq.~(\ref{5.24}) we get
\begin{equation}
\Phi_i = -2 \sqrt{-1} \partial \mbar \partial f_i, \label{5.35}
\end{equation}
where $f_i$ are real solutions of equations (\ref{5.8}). Comparing
Eqs.~(\ref{5.12c}) and (\ref{5.12e}) and using equations (\ref{5.8}),
for the pseudo--metrics $g_i$ we find the expression
\begin{equation}
g_i = 2 \partial \mbar \partial f_i dz d\mbar z = 4 k_i \exp [-(kf)_i] dz
d \mbar z. \label{5.36}
\end{equation}
Taking into account Eq.~(\ref{5.12f}) and again using equations
(\ref{5.8}), we have
\begin{equation}
R_i = - 2\sqrt{-1} \sum_{j=1}^r k_{ij} \partial \mbar \partial f_j;
\label{5.37}
\end{equation}
and, with account of Eq.~(\ref{5.35}), we come to Eq.~(\ref{5.34}).
\end{pf}

Relations (\ref{5.34}) are called the {\it generalised infinitesimal Pl\"ucker
formulas}. As we have already noted in the introduction, the validity of
these relations was conjectured in \cite{Giv89}, and proved in
\cite{Pos91}. Our proof is based on the special choice of the K\"ahler
potentials of the pseudo--metrics, and in this sense is similar to the
proof of the usual Pl\"ucker formulas \cite{GrH78}. Note also, that the
relation between the abelian Toda fields and the K\"ahler potentials
in question, has been established for the $A$--case in \cite{GM93}, and for
the other classical series ($B, C, D$) --- in \cite{GS93}, using explicit
calculations in a coordinate parametrisation of the corresponding coset
spaces.

In the conclusion, we would like to thank F. E. Burstall, J.--L. Gervais,
B. A. Khesin, M. L. Kontsevich, Yu. I. Manin and L. M. Woodward for very
useful and illuminating discussions; and Ph. A. Griffiths for interesting
communications. One of the  authors (M.V.S.)  is grateful to the
Laboratoire de Physique Th\'eorique de l'\'Ecole Normale Sup\'erieure de
Paris for a warm hospitality during his work there in 1992 and in 1993;
and the Max--Planck Institut f\"ur Mathematik in Bonn for an invitation to
visit the Institute, that was very fruitful and pleasant. This work was
partially supported by the ISF, grant RMO000.

\appendix

\section{Geometry of Complex Lie groups}

In this appendix we collect some information about complex Lie groups
which is needed in the main text; for more details see e.g.,
\cite{Bou60,Hel62,KoN63}.

Let us begin with the necessary definitions and facts about real
Lie groups. Consider an arbitrary real Lie group $G$ with the unit element
denoted $e$. Let $x \in T_e(G)$, define the vector field $X$ by
\begin{equation}
X_g \equiv L_{g*}(x), \qquad g \in G, \label{a.1}
\end{equation}
where $L_g: h \in G \to gh \in G$ is the left translation by the element
$g$. It is clear that $X$ is a left invariant vector field.
The Lie bracket of any two left invariant vector fields is a left
invariant vector field. The same is true for an arbitrary linear
combination of left invariant vector fields; thus the space of left
invariant vector fields is a Lie subalgebra of the Lie algebra of vector
fields on $G$. This Lie algebra is called the Lie algebra of the Lie group
$G$; we denote it $\goth g$.

Let $X$ be a left invariant vector field on $G$; associate with it an
element $x = X_e \in T_e(G)$. Such a correspondence is clearly a linear
mapping. Moreover, since $X$ is left invariant, then $X_e =
L_{g^{-1}}(X_g)$ for any $g \in G$. Hence, we see that different left
invariant vector fields correspond to different elements of $T_e(G)$, and
we have an isomorphism of linear spaces. This isomorphism becomes an
isomorphism of Lie algebras if we introduce in $T_e(G)$ the structure of a
Lie algebra by
\begin{equation}
[x, y] \equiv [X, Y]_e, \qquad x, y \in T_e(G), \label{a.2}
\end{equation}
where $X$ and $Y$ are left invariant vector fields defined with the help
of Eq.~(\ref{a.1}). Thus we can identify the Lie algebra $\goth g$ with
$T_e(G)$.  Below we use the latter interpretation of the Lie algebra of a
Lie group.

The adjoint representation $\Ad: g \in G \to \Ad (g)
\in \mbox{\rm GL}({\goth g})$ of the group $G$ is defined by the relation
\begin{equation}
\Ad (g) x \equiv (L_g \circ R_{g^{-1}})_* (x), \qquad g \in G,
\label{a.3}
\end{equation}
where $R_g: h \in G \to hg \in G$ is the right translation by the element
$g$. The corresponding adjoint representation $\ad: x \in {\goth g}
\to \ad (x) \in \mbox{\rm GL}({\goth g})$ of the Lie algebra ${\goth g}$ is
given by
\begin{equation}
\ad (x) y \equiv [x, y], \qquad x, y \in {\goth g}. \label{a.4}
\end{equation}

The Maurer--Cartan form of $G$ is a 1-form $\theta$ on $G$ taking values
in $\goth g$ and defined by the relation
\begin{equation}
\theta(x) \equiv L_{g^{-1}*}(x)  \label{a.5}
\end{equation}
for any $x \in T_e(G)$. The Maurer--Cartan form is left invariant, i.e.,
\begin{equation}
L_{g*} \theta = \theta \label{a.5a}
\end{equation}
for any $g \in G$. The transformation of $\theta$ under the right translations
is described by the formula
\begin{equation}
R_{g*} \theta = \Ad (g^{-1}) \circ \theta. \label{a.5b}
\end{equation}
It can be also shown that
\begin{equation}
d\theta(X, Y) + [\theta(X), \theta(Y)] = 0 \label{a.6}
\end{equation}
for any vector fields $X$ and $Y$ on $G$.

Let $\Sigma$ be a homomorphism of the group $G$, and
\begin{equation}
\sigma \equiv \Sigma_{*e} \label{a.6a}
\end{equation}
be the corresponding homomorphism of ${\goth g}$. It follows from the
definition of the adjoint representation of $G$ that
\begin{equation}
\sigma \circ \Ad (g) = \Ad (\Sigma(g)) \circ \sigma. \label{a.6b}
\end{equation}
Furthermore, it can be easily shown that
\begin{equation}
\Sigma^* \theta = \sigma \circ \theta. \label{a.6c}
\end{equation}

In this paper we mainly deal with complex differentiable
manifolds. Recall some relevant definitions and introduce
notations. Let $V$ be a complex linear space. On the space $V_{\Bbb R}$
there is a natural complex structure $J^V$ generated by the multiplication
by $\sqrt{-1}$ in $V$.  Consider the space $(V_{\Bbb R})^{\Bbb C}$.  The linear
space $V_{\Bbb R}$ can be naturally considered as a real subspace of
$(V_{\Bbb R})^{\Bbb C}$.  Any element $v$ of $(V_{\Bbb R})^{\Bbb C}$ has a
unique representation of the form
\begin{equation}
v = u + \sqrt{-1} w, \label{a.7}
\end{equation}
where $u, w \in V_{\Bbb R}$. The complex conjugation in $(V_{\Bbb
R})^{\Bbb C}$ is defined by
\begin{equation}
\mbar v = u - \sqrt{-1} w, \label{a.8}
\end{equation}
where $u, w$ are elements of $V_{\Bbb R}$ entering representation
(\ref{a.7}).

Let $W$ be another complex linear space, and $\varphi$ be a linear mapping
from $(V_{\Bbb R})^{\Bbb C}$ to $(W_{\Bbb R})^{\Bbb C}$. The complex
conjugate mapping $\mbar \varphi$ acts on the elements of $(V_{\Bbb
R})^{\Bbb C}$ according to the rule
\begin{equation}
\mbar \varphi v \equiv \mbar{ \varphi\mbar v}, \qquad v \in (V_{\Bbb
R})^{\Bbb C}. \label{a.9}
\end{equation}
The mapping $\varphi$ is called real if $\mbar \varphi = \varphi$.

The operator $J^V$ can be naturally extended to the linear operator on
$(V_{\Bbb R})^{\Bbb C}$, which will be also denoted $J^V$. Define the
subspaces $V^{(1,0)}$ and $V^{(0,1)}$ of the space $(V_{\Bbb R})^{\Bbb C}$
by
\begin{eqnarray}
V^{(1,0)} &=& \{ v \in (V_{\Bbb R})^{\Bbb C} \mid J^V v = \sqrt{-1} v \},
\label{a.10} \\
V^{(0,1)} &=& \{ v \in (V_{\Bbb R})^{\Bbb C} \mid J^V v = -\sqrt{-1}v \}.
\label{a.11}
\end{eqnarray}
The space $(V_{\Bbb R})^{\Bbb C}$ is a direct sum of the subspaces
$V^{(1,0)}$ and $V^{(0,1)}$. Hence, any element $v \in (V_{\Bbb R})^{\Bbb
C}$ can be uniquely written in the form
\begin{equation}
v = v^{(1,0)} + v^{(0,1)}, \label{a.12}
\end{equation}
where $v^{(1,0)} \in V^{(1,0)}$ and $v^{(0,1)} \in V^{(0,1)}$. The
linear operators
\begin{equation}
P^V \equiv \frac{1}{2} (I - \sqrt{-1} J^V), \qquad \mbar P^V \equiv
\frac{1}{2}(I + \sqrt{-1} J^V). \label{a.13}
\end{equation}
project $(V_{\Bbb R})^{\Bbb C}$ on the subspaces $V^{(1,0)}$ and
$V^{(0,1)}$, respectively. Note that $P^V$ and $\mbar P^V$ are
connected by the complex conjugation, that is already reflected in their
notation.

Let $M$ be a complex manifold. The natural complex structure on $M$ will
be denoted $J^M$; by $T_{{\Bbb R}p}(M)$ we denote the usual real tangent
space to $M$ at the point $p$, where $M$ is considered as a real
manifold; and the complexification of $T_{{\Bbb R}p}(M)$ is denoted by
$T^{\Bbb C}_p(M)$. As above, we have the decomposition of $T^{\Bbb
C}_p(M)$ into a direct sum
\begin{equation}
T^{\Bbb C}_p(M) = T^{(1,0)}(M) \oplus T^{(0,1)}(M). \label{a.14}
\end{equation}
Let $P^M_p$ be the mapping projecting an element of $T^{\Bbb
C}_p(M)$ to its (1,0)--component, in other words
\begin{equation}
P^M_p x \equiv \frac{1}{2} (x - \sqrt{-1} J^M_p x) \label{a.15}
\end{equation}
for any vector $x \in T^{\Bbb C}_p(M)$. The mapping
$P^M$ projecting a vector field $X$ on $M$ to its $(1,0)$--component is
defined by
\begin{equation}
(P^M X)_p \equiv P^M_p X_p \label{a.16}
\end{equation}
for any vector field $X$ on $M$. The projection mappings to the
(0,1)--com\-po\-nent of a tangent vector and a vector field are defined
similarly.

Now, let $G$ be a complex Lie group. In this case the group operation
$(g,h) \in G \times G \to gh^{-1} \in G$ is, by definition, a holomorphic
mapping.  The group $G$ can be considered as a real Lie group
supplied with a complex structure $J^G$.  Denote the corresponding
real Lie algebra ${\goth g}_{\Bbb R}$.  Since the group operation in $G$ is
holomorphic, we have
\begin{eqnarray}
L_{g*} \circ J^G &=& J^G \circ L_{g*}, \label{a.17} \\
R_{g*} \circ J^G &=& J^G \circ R_{g*} \label{a.18}.
\end{eqnarray}

The restriction of the complex structure $J^G$ to $T_e(G)$ generates a
complex structure $J^{{\goth g}_{\Bbb R}}$ on the Lie algebra ${\goth
g}_{\Bbb R}$. It follows from Eqs.~(\ref{a.17}) and (\ref{a.18}) that
\begin{equation}
\ad (x) \circ J^{{\goth g}_{\Bbb R}} = J^{{\goth g}_{\Bbb R}} \circ
\ad (x) \label{a.19}
\end{equation}
for any $x \in {\goth g}_{\Bbb R}$. Hence, $J^{{\goth g}_{\Bbb R}}$ is a
Lie complex structure, and ${\goth g}_{\Bbb R}$ has the structure of a
complex Lie algebra; denote it $\goth g$. Actually the real Lie algebra
${\goth g}_{\Bbb R}$ is the realification of the complex Lie algebra
${\goth g}$, that is already reflected in the notation ${\goth g}_{\Bbb
R}$.  The (complex) Lie algebra ${\goth g}$ is called the Lie algebra of
the (complex) Lie group $G$. In fact, there exists another interpretation
of the Lie algebra of a complex Lie group, that is more convenient for our
purposes. To give this interpretation let us recall some facts from the
theory of Lie algebras.

Let $\goth g$ be a complex Lie algebra.  Consider the space $({\goth
g}_{\Bbb R})^{\Bbb C}$. This space has a natural structure of a complex
Lie algebra.  It can be shown that the decomposition
\begin{equation}
({\goth g}_{\Bbb R})^{\Bbb C} = {\goth g}^{(1,0)} \oplus {\goth g}^{(0,1)}.
\label{a.20}
\end{equation}
is a decomposition of $({\goth g}_{\Bbb R})^{\Bbb C}$ into a direct sum
of two Lie subalgebras.
Moreover, we have
\begin{eqnarray}
&&P^{\goth g} (ax + bJ^{{\goth g}_{\Bbb R}} x) = (a + \sqrt{-1} b)
P^{\goth g} x, \qquad P^{\goth g} [x, y] = [P^{\goth g} x, P^{\goth
g} y], \label{a.21} \\
&& \mbar P^{\goth g}(ax - bJ^{{\goth g}_{\Bbb R}} x) = (a + \sqrt{-1}
b) \mbar P^{\goth g} x, \qquad \mbar P^{\goth g} [x, y] = [\mbar
P^{\goth g} x, \mbar P^{\goth g} y] \label{a.22}
\end{eqnarray}
for all $a, b \in {\Bbb R}$ and $x, y \in ({\goth g}_{\Bbb R})^{\Bbb C}$.
Restricting $P^{\goth g}$ and $\mbar P^{\goth g}$ to ${\goth g}_{\Bbb R}$,
we get bijective mappings from ${\goth g}_{\Bbb R}$ to ${\goth g}^{(1,0)}$
and ${\goth g}^{(0,1)}$. Taking into account Eqs.~(\ref{a.21}) and
(\ref{a.22}), we see that the first mapping sets an isomorphism between
the Lie algebras ${\goth g}$ and ${\goth g}^{(1,0)}$, while the second one
sets an isomorphism between the Lie algebras $\mbar {\goth g}$ and ${\goth
g}^{(0,1)}$. Here $\mbar{\goth g}$ is a complex Lie algebra which is
obtained from ${\goth g}_{\Bbb R}$ with the help of the Lie complex
structure $-J^{{\goth g}_{\Bbb R}}$.

Thus, we can identify the Lie algebra of the Lie group $G$ with the
holomorphic tangent space $T^{(1,0)}_e(G) \subset T^{\Bbb C}_e(G)$.
Actually, any element of the space $T^{(1,0)}_e(G)$ generates a holomorphic
left invariant vector field on $G$ given by the relation of form
(\ref{a.1}); and the Lie algebra operation in $\goth g$ is related to the
Lie bracket of the corresponding holomorphic left invariant vector fields
by Eq.~(\ref{a.2}). Note that any left invariant vector field on $G$ of
the type (1,0) is a holomorphic vector field.

Considering $G$ as a real Lie group, we can define on it the corresponding
Maurer--Cartan form $\theta_{\Bbb R}$ taking values in ${\goth g}_{\Bbb
R}$. It is clear that
\begin{equation}
\theta_{\Bbb R}(J^G(X)) = J^G_e \circ \theta_{\Bbb R}(X) \label{a.23}
\end{equation}
for any vector field $X$ on $G_{\Bbb R}$.  The complexification
$(\theta_{\Bbb R})^{\Bbb C}$ takes values in $({\goth g}_{\Bbb R})^{\Bbb
C}$. The Maurer--Cartan form $\theta$ of the complex Lie group $G$ is
defined by
\begin{equation}
\theta = P^G_e \circ (\theta_{\Bbb R})^{\Bbb C}. \label{a.24}
\end{equation}
It follows from this definition that $\theta$ takes values in the Lie
algebra ${\goth g}$.  Using Eq.~(\ref{a.23}), we get
\begin{equation}
\theta(X) = (\theta_{\Bbb R})^{\Bbb C}(P^G X); \label{a.25}
\end{equation}
hence, $\theta$ is a 1-form of the type (1,0). Moreover, it can be shown
that it is a holomorphic 1-form.

Actually, we can define $\theta$ as the unique 1-form of the type (1,0),
satisfying relation (\ref{a.5}), where $X$ now is an arbitrary left
invariant vector field on $G$ of the type (1,0). It can be shown that in
the case of a complex Lie group we still have relations
(\ref{a.5a})--(\ref{a.6}). The situation with relation (\ref{a.6c}) is
more complicated. In fact, it is natural to consider only holomorphic and
antiholomorphic automorphisms of $G$. For a holomorphic automorphism
$\Sigma$ we have that Eq.~(\ref{a.6c}) is valid with $\sigma$ defined by
Eq.~(\ref{a.6a}). If $\Sigma$ is an antiholomorphic automorphism of $G$,
then
\begin{equation}
\Sigma_{*e} T^{(1,0)}_e(G) = T^{(0,1)}_e(G). \label{a.28}
\end{equation}
In this case we define $\sigma$ as an antilinear automorphism of ${\goth
g}$ by
\begin{equation}
\sigma x \equiv \mbar{\Sigma_{*e}(x)} \qquad x \in {\goth g}. \label{a.29}
\end{equation}
This definition leads to the equality
\begin{equation}
\Sigma^* \theta(x) = (\sigma \circ \theta)(\mbar x), \label{a.30}
\end{equation}
which is valid for any $x \in T^{\Bbb C}_g(G)$, $g \in G$.
Note that relation (\ref{a.6b}) is valid for both cases with the mapping
$\sigma$ defined either by Eq.(\ref{a.6a}), or by Eq.(\ref{a.29}).

\section{Complex Lie Algebras, Flag Manifolds, and ${\Bbb Z}$--Gradations}

In this appendix we present some facts on complex semisimple finite
dimensional Lie algebras, see e.g. \cite{Bou60,GoG78,Hel62,GOV93}, needed
in the main text. We also consider here the concept of a flag manifold and
its relation to the concept of ${\Bbb Z}$--gradation of a Lie algebra.

Let ${\goth g}$ be a complex semisimple Lie algebra, and ${\goth h}$ be a
fixed Cartan subalgebra of ${\goth g}$. Further, let $\Delta$
be the system of the nonzero roots of ${\goth g}$ with respect to ${\goth h}$.
The root subspaces ${\goth g}^\alpha$, $\alpha \in \Delta$, are defined as
\begin{equation}
{\goth g}^\alpha \equiv \{ x \in {\goth g} \mid [h,\,x] = \alpha(h) x
\mbox{ for all } h \in {\goth h}\}. \label{b.1a}
\end{equation}
All the root subspaces are one--dimensional, and the Lie algebra ${\goth
g}$ is represented as a direct sum
\begin{equation}
{\goth g} = {\goth h} \oplus \bigoplus_{\alpha \in \Delta} {\goth
g}^\alpha. \label{b.1b}
\end{equation}
Choose in each subspace ${\goth g}^\alpha$ a nonzero vector $x_\alpha$,
then
\begin{equation}
[x_\alpha, x_\beta] = n_{\alpha, \beta} x_{\alpha + \beta}, \label{b.1}
\end{equation}
if $\alpha + \beta \in \Delta$, while
\begin{equation}
[x_\alpha, x_\beta] = 0, \label{b.2}
\end{equation}
if $\alpha + \beta \not\in \Delta$ and $\alpha + \beta \ne 0$.

Let $(\ ,\ )$ be the Killing--Cartan form of ${\goth g}$. Define
the vectors $h_\alpha \in {\goth h}$,  $\alpha \in \Delta$, by
\begin{equation}
(h, h_\alpha) = \alpha(h), \qquad h \in {\goth h}. \label{b.3}
\end{equation}
It can be shown that the vectors $x_\alpha$ can be chosen in
such a way that
\begin{equation}
[x_\alpha, x_{-\alpha}] = h_\alpha, \label{b.4}
\end{equation}
and the constants $n_{\alpha, \beta}$ satisfy the relation
\begin{equation}
n_{\alpha, \beta} = - n_{-\alpha, -\beta}. \label{b.5}
\end{equation}
In this case $n_{\alpha, \beta}$ are real numbers. The set
of the $x_\alpha$, $\alpha \in \Delta$, having the properties described
above, is called the Weyl basis of ${\goth g}$ modulo ${\goth h}$.

Recall that any involutive antilinear automorphism $\sigma$ of
${\goth g}$ specifies a real form $\tilde {\goth g}$ of ${\goth g}$
defined by
\begin{equation}
{\goth g}_\sigma \equiv \{ x \in {\goth g} \mid \sigma(x) = x\}.
\label{b.6}
\end{equation}
Using the Weyl basis of ${\goth g}$, we can define an antilinear
automorphism $\sigma$ of ${\goth g}$ with the help of the relations
\begin{equation}
\sigma(h_\alpha) = - h_\alpha, \qquad \sigma(x_\alpha) = -x_{-\alpha}.
\label{b.7}
\end{equation}
The corresponding real form ${\goth u}$ in this case is compact.

Let $\Delta = \Delta_- \cup \Delta_+$ be a representation of the root
system of ${\goth g}$ as a disjoint union of the positive and negative
roots with respect to some lexicographic ordering, and $\Pi
\subset \Delta_+$ be a system of simple roots. Denote the elements of
$\Pi$ by $\alpha_i$, $i = 1,\ldots, r \equiv \rank {\goth g}$. Let
$\{x_\alpha\}$ be the Weyl basis of ${\goth g}$. The elements
\begin{equation}
h_i \equiv \frac{2}{(\alpha_i, \alpha_i)} h_{\alpha_i} \qquad x_{\pm i}
\equiv x_{\pm\alpha_i} \label{b.8}
\end{equation}
generate the whole Lie algebra ${\goth g}$. Under an appropriate
normalisation, the elements $x_{\pm i}$, together with the Cartan
generators $h_i$, satisfy the defining relations
\begin{equation}
[h_i, h_j] = 0, \qquad [h_i, x_{\pm j}] = \pm k_{ji} x_{\pm j}, \qquad
[x_{+i}, x_{-j}] = \delta_{ij} h_i, \label{b.9}
\end{equation}
where $k$ is the Cartan matrix of ${\goth g}$. The elements $x_{\pm i}$
are called Chevalley generators of ${\goth g}$.

A Lie subalgebra ${\goth b}$ is called a Borel subalgebra of
${\goth g}$ if ${\goth b}$ is a maximal solvable subalgebra of ${\goth
g}$.  Denote by $G$ a complex connected Lie group having ${\goth g}$ as
its Lie algebra. A subgroup $B$ of $G$ is called a Borel subgroup if its
Lie algebra ${\goth b}$ is a Borel subalgebra of ${\goth g}$.  According
to the Borel--Morozov theorem, all Borel subalgebras of ${\goth g}$ are
connected one to each other by the action of the group $\Aut\,{\goth
g}$ of automorphisms of ${\goth g}$. The similar statement is valid also
for the Borel subgroups of $G$.

The root decomposition (\ref{b.1b}) of ${\goth g}$ implies that
\begin{equation}
{\goth g} = {\goth n}_- \oplus {\goth h} \oplus {\goth n}_+, \label{b.10}
\end{equation}
where
\begin{equation}
{\goth n_\pm} \equiv \bigoplus_{\alpha \in \Delta_\pm} {\goth g}^\alpha
\label{b.11}
\end{equation}
are nilpotent subalgebras of ${\goth g}$. The corresponding connected
nilpotent Lie subgroups of $G$ will be denoted $N_\pm$. It can be shown
that the subalgebras
\begin{equation}
{\goth b}_\pm \equiv {\goth h} \oplus {\goth n}_\pm \label{b.12}
\end{equation}
are Borel subalgebras of ${\goth g}$. The Borel subalgebra ${\goth b}_-$
is called opposite to the Borel subalgebra ${\goth b}_+$. The
subalgebras ${\goth b_\pm}$ generate Borel subgroups denoted $B_\pm$.

A parabolic subalgebra of ${\goth g}$ is, by definition, a subalgebra of
${\goth g}$ which contains a Borel subalgebra of ${\goth g}$. A subgroup
$P$ of the Lie group $G$ is called a parabolic subgroup if it contains
a Borel subgroup of $G$.  Any subsystem $\Psi$ of the system of simple
roots $\Pi$ generates parabolic subalgebras
\begin{equation}
{\goth p}_{\pm \Psi} \equiv {\goth b}_{\pm} \oplus \bigoplus_{\alpha \in
[\Psi] \cap \Sigma_\mp} {\goth g}^\alpha, \label{b.13}
\end{equation}
where $[\Psi]$ is the set of all roots from $\Delta$ which can be
represented as linear combinations of the roots from $\Psi$. It can be
shown that, up to transformations of the group $\Aut \,{\goth g}$, any
parabolic subalgebra of ${\goth g}$ can be obtained in such a way. Thus,
we have a transparent classification of the parabolic subalgebras of any
complex semisimple Lie algebra.  One usually writes ${\goth p}_{\pm \{i_1,
\ldots, i_k\}}$ for the parabolic subalgebra corresponding to the
subsystem $\Psi = \{\alpha_{i_1}, \ldots, \alpha_{i_k}\}$. The parabolic
subalgebras corresponding to a subsystem which consists of just one simple
root $\alpha_i$, is denoted ${\goth p}_{\pm i}$. At last, the parabolic
subalgebras corresponding to a subsystem consisting of all simple roots
except $\alpha_i$, is denoted ${\goth p}_{\pm i'}$.

Let $P$ be a parabolic subgroup of $G$.  The homogeneous space $F \equiv
G/P$ is called a flag manifold or, quite rarely, a parabolic space. It is
clear that $F$ has the natural structure of a complex manifold. Let
${\goth u}$ be a compact form of ${\goth g}$ defined by the antilinear
involutive automorphism $\sigma$ given by relations (\ref{b.7}), and
${\goth a}$ be a maximal abelian subalgebra of ${\goth u}$.  Denote by $U$
and $A$ the corresponding real connected Lie groups.  According to the
Borel--Hirzebruch theorem, the flag manifolds $G/B_\pm$ are diffeomorphic
to the homogeneous space $U/A$. It follows from this fact that for any
Borel subgroup $B$, the flag manifold $G/B$ is compact. Since any parabolic
subgroup $P$ contains some Borel subgroup $B$, there is the natural
projection from $G/B$ to $G/P$; hence all flag manifolds are compact.

Consider the realification ${\goth g}_{\Bbb R}$ of the Lie algebra ${\goth
g}$. Let $J^{{\goth g}_{\Bbb R}}$ be the linear operator in ${\goth
g}_{\Bbb R}$, corresponding to the multiplication by $\sqrt{-1}$ in
${\goth g}$.  The expansion of ${\goth g}_{\Bbb R}$ into the direct sum
\begin{equation}
{\goth g}_{\Bbb R} = {\goth u} \oplus J^{{\goth g}_{\Bbb R}} {\goth a}
\oplus {\goth n}_{+\Bbb R}, \label{b.14}
\end{equation}
where subalgebras ${\goth u}$, ${\goth a}$ and ${\goth n}_+$ were defined
above, is called the Iwasawa decomposition of ${\goth g}$. Note that
\begin{equation}
{\goth h}_{\Bbb R} = {\goth a} \oplus J^{{\goth g}_{\Bbb R}}{\goth a}.
\label{b.14a}
\end{equation}
There is the corresponding analytic decomposition of the Lie group $G$
considered as a real Lie group. It has the form
\begin{equation}
G = U N_+ A^*, \label{b.15}
\end{equation}
where $A^*$ is the real connected Lie group corresponding to the
subalgebra $J^{{\goth g}_{\Bbb R}} {\goth a}$.

A decomposition of ${\goth g}$ into a direct sum of subspaces ${\goth
g}_m$,
\begin{equation}
{\goth g} = \bigoplus_{m \in {\Bbb Z}} {\goth g}_m, \label{b.16}
\end{equation}
such that
\begin{equation}
[{\goth g}_m, {\goth g}_k] \subset {\goth g}_{m+k}, \label{b.17}
\end{equation}
is called a ${\Bbb Z}$--gradation of ${\goth g}$. For any ${\Bbb
Z}$--gradation of ${\goth g}$ we can construct a homomorphism $\rho:
{\Bbb C}^* \to \Aut \,{\goth g}$, where ${\Bbb C}^* \equiv {\Bbb C} -
\{0\}$ is considered as a complex Lie group with respect to the
multiplication. The homomorphism $\rho$ is defined by
\begin{equation}
\rho(\lambda) x = \rho(\lambda)\left(\sum_{m \in {\Bbb Z}} x_m
\right) \equiv \sum_{m \in {\Bbb Z}} \lambda^m x_m, \qquad \lambda \in
{\Bbb C}^*. \label{b.18}
\end{equation}
Actually, the mapping $\rho$ takes values in the connected component of the
group $\Aut \,{\goth g}$ containing unity, which, for the considered
case of a semisimple Lie algebra, coincides with the Lie group
$\Int \,{\goth g}$ of internal automorphisms of ${\goth g}$. The Lie algebra
of the group $\Int \,{\goth g}$ is the Lie algebra $\ad \,{\goth
g}$. Hence, there exists an element $q$ of ${\goth g}$ such that
\begin{equation}
[q, x_m] = m x_m \label{b.19}
\end{equation}
for any $x_m \in {\goth g}_m$. The element $q$ is called the gradation
operator of a given ${\Bbb Z}$--gradation. Thus, for a complex semisimple
Lie algebra, any ${\Bbb Z}$--gradation may be defined with the help of the
corresponding gradation operator.

It is clear that the linear operator $\ad (q)$ is semisimple and
satisfies the relation
\begin{equation}
\exp(2\pi i\,\ad (q)) = \id_{\goth g}. \label{b.20}
\end{equation}
Since we consider the case of a semisimple Lie algebra, the element $q$
is semisimple. From the other hand, it is clear that any semisimple
element $q$ of ${\goth g}$ satisfying Eq.~(\ref{b.20}), can be considered
as the gradation operator of some ${\Bbb Z}$--gradation of ${\goth g}$.

Since the element $q$ is semisimple, without any loss of generality we can
suppose that $q \in {\goth h}$ and $\alpha_i(q) \ge 0$, $i = 1, \ldots, r$.
It follows from  Eq.~(\ref{b.20}) that for any $i$, the number $n_i \equiv
\alpha_i(q)$ is an integer. The element $q$ is uniquely defined by the
numbers $n_i$. Indeed, from Eq.~(\ref{b.9}) we have
\begin{equation}
q = \sum_{i,j = 1}^r k^{-1}_{ij} n_j h_i. \label{b.21}
\end{equation}

Note that the subspace ${\goth g}_m$ is the sum of the root spaces
${\goth g}_\alpha$, corresponding to the roots $\alpha = \displaystyle
\sum_{1 \le i \le r} c_i \alpha_i$ with $\displaystyle\sum_{1 \le i \le
r} c_i n_i = m$. The subspace ${\goth g}_0$ includes also the Cartan
subalgebra ${\goth h}$. Any positive root belongs to some
subspace ${\goth g}_m$ with $m \ge 0$, while any negative root belongs to
some subspace ${\goth g}_m$ with $m \le 0$. Further, if ${\goth g}^\alpha
\subset {\goth g}_m$, then ${\goth g}^{-\alpha} \subset {\goth g}_{-m}$;
hence, we have
\begin{equation}
\dim \,{\goth g}_m = \dim \,{\goth g}_{-m}. \label{b.37}
\end{equation}

The subspaces $\tilde {\goth n}_\pm$, defined by
\begin{equation}
\tilde {\goth n}_- \equiv \bigoplus_{m < 0} {\goth g}_m, \qquad
\tilde {\goth n}_+ \equiv \bigoplus_{m > 0} {\goth g}_m. \label{b.22}
\end{equation}
are nilpotent subalgebras of ${\goth g}$. It is clear that
\begin{equation}
\tilde {\goth n}_\pm \subset {\goth n}_\pm. \label{b.23}
\end{equation}
Introduce the notation
\begin{equation}
\tilde {\goth b}_- \equiv \bigoplus_{m \le 0} {\goth g}_m, \qquad
\tilde {\goth b}_+ \equiv \bigoplus_{m \ge 0} {\goth g}_m \label{b.24}
\end{equation}
Since ${\goth h} \subset {\goth g}_0$, we conclude that $\tilde {\goth
b}_\pm \supset {\goth b}_\pm$. Hence $\tilde {\goth b}_\pm$ are
parabolic subalgebras of ${\goth g}$. Defining the set $\Pi_0 \subset \Pi$
by
\begin{equation}
\Pi_0 \equiv \{ \alpha_i \in \Pi \mid \alpha_i(q) = 0 \}, \label{b.25}
\end{equation}
we can write
\begin{equation}
\tilde {\goth b}_\pm = {\goth p}_{\pm \Pi_0}. \label{b.26}
\end{equation}
{}From the other hand, let a parabolic subalgebra ${\goth p}_\Psi$ with
$\Psi$ being a subsystem of $\Pi$, be given. Consider an element $q \in
{\goth h}$, such that $\alpha_i(q) = 0$ if and only if $\alpha_i \in
\Psi$, and all numbers $\alpha_i(q)$ are nonnegative integers. It is clear
that the element $q$ is the gradation operator for a ${\Bbb Z}$--gradation
of ${\goth g}$ with
\begin{equation}
\tilde {\goth b}_\pm = {\goth p}_{\pm \Psi}. \label{b.27}
\end{equation}
Thus, for any parabolic subalgebra ${\goth p}$ of ${\goth g}$ we can find
a ${\Bbb Z}$--gradation of ${\goth g}$, such that the corresponding
subalgebra $\tilde {\goth p}_+$ (or $\tilde {\goth p}_-$)
is connected with ${\goth p}$ by a transformation of the group
$\Aut \,{\goth g}$.

It is natural to denote the subalgebra ${\goth g}_0$ by $\tilde {\goth
h}$. Recall that $\tilde {\goth h}$ is not, in general, a Cartan
subalgebra of ${\goth g}$, but it contains some such subalgebra. Using
the notations introduced above, we can write
\begin{equation}
\tilde {\goth b}_\pm = \tilde {\goth h} \oplus \tilde {\goth
n}_\pm, \label{b.28}
\end{equation}
and
\begin{equation}
{\goth g} = \tilde {\goth n}_- \oplus \tilde {\goth h}
\oplus \tilde {\goth n}_+. \label{b.29}
\end{equation}

For any complex semisimple Lie algebra there is one distinguished ${\Bbb
Z}$--gradation, arising when we choose all the numbers $n_i$ equal to 1.
The corresponding gradation operator have in this case the form
\begin{equation}
q = \sum_{i = 1}^r k_i h_i, \label{b.30}
\end{equation}
where
\begin{equation}
k_i \equiv \sum_{j=1}^r k^{-1}_{ij}. \label{b.31}
\end{equation}
Such a gradation is called the canonical gradation. For the canonical
gradation all the quantities with tildes defined above are equal to the
corresponding quantities without tildes; and, of course, the subalgebra
${\goth g}_0$ is abelian; it is a Cartan subalgebra.  Note also that in
this case the subspaces ${\goth g}_{\pm 1}$ coincide with the linear spans
of the vectors $x_{\pm i}$, $i = 1,\ldots, r$.

It is convenient for our purposes to consider ${\Bbb Z}$--gradations of
${\goth g}$ related to the embeddings of the Lie algebra $\mbox{\goth sl}(2,
{\Bbb C})$ into ${\goth g}$. Recall that the Lie algebra $\mbox{\goth sl}(2,
{\Bbb C})$ consists of all traceless $2\times 2$ matrices. The matrices
\begin{equation}
x_- \equiv \left(\begin{array}{cc}
0 & 0 \\
1 & 0
\end{array} \right), \qquad
h \equiv \left(\begin{array}{cr}
1 & 0 \\
0 & -1
\end{array} \right), \qquad
x_+ \equiv \left(\begin{array}{cc}
0 & 1 \\
0 & 0
\end{array} \right) \label{b.32}
\end{equation}
form a basis in $\mbox{\goth sl}(2, {\Bbb C})$ called the canonical basis.
For these matrices we have
\begin{equation}
[h, x_\pm] = \pm 2 x_\pm, \qquad [x_+, x_-] = h. \label{b.33}
\end{equation}
By an embedding of $\mbox{\goth sl}(2, {\Bbb C})$ into ${\goth g}$ we mean a
nontrivial homomorphism from $\mbox{\goth sl}(2, {\Bbb C})$ into ${\goth g}$;
their complete classification has been done by E. B. Dynkin.
The images of the elements $h$ and $x_\pm$ under such a homomorphism are
denoted usually by the same letters.  From the properties of the
representations of the Lie algebra $\mbox{\goth sl}(2, \Bbb C)$ it follows,
that the element $h$ of ${\goth g}$ should be semisimple, and the elements
$x_\pm \in {\goth g}$ should be nilpotent. Moreover, it is clear that
\begin{equation}
\exp(2\pi i\,\ad (h)) = \id_{\goth g}. \label{b.34}
\end{equation}
As above, without any loose of generality, one can suppose that $h \in {\goth
h}$, and $\alpha_i(h) \ge 0$ for all $\alpha_i \in \Pi$. It can be
shown that the numbers $\alpha_i(h)$ are equal only to 0, 1 and 2. So, the
element $h$ can be considered as the gradation operator of some ${\Bbb
Z}$--gradation of ${\goth h}$. In fact, it is more convenient for our
purposes to define the gradation operator $q$, connected with the given
embedding of $\mbox{\goth sl}(2, {\Bbb C})$ into ${\goth g}$, by the relation
\begin{equation}
h = 2q. \label{b.35}
\end{equation}
It is clear that this definition leads to the necessity to consider also
${\Bbb Z}/2$--gradations of ${\goth g}$. If, instead of Eq.~(\ref{b.34}), we
have the relation
\begin{equation}
\exp(\pi i\,\ad (h)) = \id_{\goth g}, \label{b.36}
\end{equation}
we call the corresponding embedding integral, otherwise we deal with a
semi--integral embedding. For an integral embedding the numbers $\alpha_i(h)$
are equal to 0 or 2.

The properties of the representations of the algebra $\mbox{\goth sl}(2,
{\Bbb C})$ imply that if we consider a ${\Bbb Z}$--gradation, or a ${\Bbb
Z}/2$--gradation, associated with an embedding of $\mbox{\goth sl}(2,
{\Bbb C})$ into ${\goth g}$, then
\begin{equation}
\dim \,{\goth g}_0 \ge \dim \,{\goth g}_{\pm 1}. \label{b.38}
\end{equation}

For any complex semisimlpe Lie algebra there is an embedding of
$\mbox{\goth sl}(2, {\Bbb C})$ leading to the canonical gradation. This
embedding is defined by
\begin{equation}
x_{\pm} = \sum_{i=1}^r (2k_i)^{1/2} x_{\pm i}, \qquad h = 2 \sum_{i=1}^r
k_i h_i, \label{b.39}
\end{equation}
with $k_i$, given by Eq.~(\ref{b.31}), and is called the principal embedding;
its exhaustive investigation has been done by B. Kostant.

It is known that not any element $h \in {\goth h}$, even satisfying the
requirement $\alpha_i(h) = 0, 1$ or 2, can be considered as the
corresponding element of some embedding of $\mbox{\goth sl}(2, {\Bbb C})$ into
${\goth g}$, and we have not here a direct relation to parabolic
subalgebras, as it was for the case of a general ${\Bbb Z}$--gradation.

\end{document}